\documentclass[manuscript]{aastex}

\usepackage{amsmath, amssymb}

\shorttitle{Modulation mechanism in LS5039}
\shortauthors{Yamaguchi \& Takahara 2010}

\begin{document}


\title{Modulation Mechanism of TeV, GeV and X-ray Emission in LS5039}


\author{M. S. Yamaguchi and F. Takahara}
\affil{Department of Earth and Space Science, Graduate School of Science, Osaka University,
    Toyonaka Osaka, 560-0043, Japan}




\begin{abstract}
The emission mechanism of the gamma-ray binary LS5039 in energy bands of TeV, GeV, and X-ray is investigated.
Observed light curves in LS5039 show that TeV and GeV fluxes anticorrelate and TeV and X-ray fluxes correlate. 
However, such correlated variations have not been explained yet reasonably at this stage. 
Assuming that relativistic electrons are injected constantly at the location of the compact object as a point source, 
and that they lose energy only by the inverse Compton (IC) process, we calculate gamma-ray spectra and light curves 
by the Monte Carlo method, including the full electromagnetic cascade process. Moreover, we calculated X-ray spectra and 
light curves by using the resultant electron distribution. 
As a result, we are able to reproduce qualitatively spectra and light curves observed by \textit{HESS}, \textit{Fermi}, and \textit{Suzaku} for the
inclination angle $i = 30^{\circ}$ and the index of injected electron distribution $p = 2.5$. 
We conclude that TeV-GeV anticorrelation is due to anisotropic IC scattering and anisotropic 
$\gamma \gamma$ absorption, and that TeV-X correlation is due to the dependence of IC cooling time on orbital phases.
In addition, the constraint on the inclination angle implies that the compact object in LS5039 is a black hole.
\end{abstract}





\section{Introduction}
\label{intro}
Three gamma-ray binaries have been found so far, which consist of a massive star and a compact object (a neutron star 
or a black hole). They are LS5039 \citep{aha05a}, LSI +61$^{\circ}$ 303 \citep{alb06}, and PSR B1259-63 \citep{aha05b}. 
The compact objects of the first two have not yet been identified, while that of the last one is known to be a young pulsar. 

It is anisotropic IC process and $\gamma \gamma$ absorption that are the important electromagnetic processes for high energy electrons, positrons and gamma-rays occurring 
in gamma-ray binaries \citep[e.g.][]{dub06a,bed06,dub08,kha08}.
Because of thick radiation field around the OB star, electrons and positrons which are accelerated near the compact 
object radiate TeV and GeV gamma-rays through IC process. 
Very high energy gamma-rays are absorbed by ultraviolet photons which come from the OB star, and produce $e^{\pm}$ pairs. 
Thus, if electrons and positrons lose energy by IC process, electromagnetic cascade process develops because the secondary pairs radiate a plenty of photons with enough energy to create $e^{\pm}$ pairs. 
It is also important to consider anisotropy of IC process in the binary system.
The flux of IC process, which is affected according to anisotropic radiation field, depends on the angle between 
the line of sight and the line joining the center of the OB star and the compact object, so that it depends on orbital phases.

Magnetic field affects $e^{\pm}$ propagation and radiation. If magnetic field in the system is sufficiently weak,
relativistic electrons and positrons proceed straight in a direction of injection, and radiate IC photon in the same direction. 
However, if it is strong and turbulent,
then relativistic electrons and positrons are trapped by such magnetic field, and can radiate IC photon in any direction.
It should be noted that even if synchrotron cooling is negligible compared with IC cooling, electrons and positrons can radiate synchrotron photons.
If magnetic field is sufficiently strong, they lose energy by synchrotron process.

LS5039 consists of a compact object and a massive star whose spectral type is O6.5 V and whose mass is $M_* = 22.9^{+3.4}_{-2.9}\ M_{\odot}$, 
and its binary period is 3.906 days
\citep{cas05}. In addition, its separation changes from $\sim 2.2\ R_*$ at periastron to $ \sim 4.5\ R_*$ at apastron, 
where $R_* = 9.3^{+0.7}_{-0.6}\ R_{\odot}$ represents the radius of the companion O star \citep{cas05}.

For LS5039, spectra and light curves of TeV gamma-ray with HESS array of atmospheric Cherenkov telescopes \citep{aha06a},
very recently, those of GeV gamma-ray with Fermi Gamma-ray Space Telescopes \citep{abd09}, and those of X-ray with Suzaku 
\citep{tak09} have already been reported. 
These observations show that these fluxes clearly modulate with its binary period, that GeV gamma-ray flux 
anticorrelates with TeV gamma-ray flux, and that X-ray flux correlates with TeV flux.

Two kinds of models have been suggested so far for explaining $\gamma $-ray spectra and light curves emerging from LS5039.
The first kind of models is the microquasar type model \citep[e.g.][]{bos04,par06,bed06,bed07,der06,kha08}.
Assuming that the compact object in LS5039 is a black hole, they computed the propagation of photons from the injection site in 
the jet which is located at a certain distance (0 or finite value) from the base of jet.
The other kind of model is the pulsar type model \citep[e.g.][]{dub06b,sie07,sie08a,sie08b,dub08,cer08,cer09}. 
Assuming that the compact object is a neutron star, they computed the propagation of photons 
emerging from relativistic electrons accelerated well near the neutron star, namely at the termination shock which is formed
by a collision between a pulsar wind and a stellar wind, or near the pulsar magnetosphere. 
In the papers mentioned above, \citet{bed06,bed07}, \citet{sie07,sie08a,sie08b}, and
\citet{cer09} took into account electromagnetic cascade process.

The TeV flux from LS5039 was investigated in detail by \citet{kha08}. Assuming that electrons were injected at a certain point along the jet and
the counter-jet, they calculated spectra and light curves of TeV flux taking into account anisotropic IC scattering, $\gamma \gamma$ absorption,
the advection of electrons by the jet flow, and synchrotron cooling of electrons. As a result, they reproduced well the observed spectra with HESS, 
while they were not able to reproduce the light curves.

The anticorrelation of GeV and TeV fluxes was explained by \citet{bed06}. Assuming that electrons were injected in the jet and that they cooled
only by IC process, the author performed Monte Carlo simulations for LS5039 and LSI +61$^{\circ}$ 303 taking into account anisotropic IC scattering 
and the cascade process.
Though he showed that GeV and TeV fluxes anticorrelated, his model did not match the observational data in terms of the ratio of TeV flux to GeV one.

The explanation of X-ray spectra and light curves observed by Suzaku has already been tried by \citet{tak09}.
They suggested an adiabatic cooling model, so that they explained X-ray flux modulation qualitatively and reproduced spectral index 
observed with Suzaku. When electrons lose energy adiabatically, resultant power-law index of electron distribution is
the same as the injected one unlike radiative cooling, so that observed spectral index $\Gamma \sim 1.5$ is consistent with 
synchrotron emission with electrons which are injected with $p \sim 2 $ ($N_{e, \text{inj}}(E_e) \propto E_e^{-p}$).
However, X-ray emission mechanism, especially its modulation, is not investigated in detail.

We perform Monte Carlo simulations of photon propagation based on a simple geometrical model for LS5039 and aim to understand the physical mechanism
in the system by reproducing the observational data of TeV, GeV, and X-ray energy ranges.
We assume that electrons and positrons lose energy only by IC process, so that the cascade processes develop as already discussed in the second paragraph of this section,
and we calculate the TeV, GeV gamma-ray, and X-ray spectra emerging to the observer from the system for LS5039.
Our model and calculational method are presented in detail in Section \ref{model} and the numerical results are discussed in Section
\ref{result}, where the comparison of the results with the observational data is also made. We summarize the emission mechanism in LS5039
in Section \ref{sum}.

\section{Model and Method}
\label{model}
Our model is shown as a schematic picture in Figure \ref{modpic}. The emission process we consider is shown in the left panel in Figure \ref{modpic}
and when comparing with the observational data, we determine the direction of the observer as shown in the right panel in Figure \ref{modpic}.
First of all, electrons and/or positrons are injected isotropically 
at the location of the compact object. They cool only by IC process as stated Section \ref{intro}, and radiate high energy photons
in various directions in 3-dimensional space, when they interact with thermal radiation from the companion star. 
The high energy photons also interact with photons from the companion and create $e^{\pm}$ pairs. We ignore the propagation of these 
electrons and positrons, so that they radiate IC photons at the created (and injected) places. Thus, these electrons and positrons initiate
the cascade processes.

We count photons which undergo such a cascade process and escape from the system in any direction.
By rotating the direction of the observer, we can obtain spectra by photons escaping in the direction of a certain range of azimuthal angle (the right panel of Figure \ref{modpic}). These are equivalent to spectra by escaped photons during a relevant range of the orbital phase (hereafter we call the spectra "phase-divided spectra"). At the same time, we can obtain the light curves in GeV-TeV energy range in any inclination angle, when we assume that the emission is stationary during each divided phase.
We also calculate spectra of synchrotron radiation by electrons and positrons in the system, assuming uniform magnetic field and using 
the electron distribution in steady state derived by cascade process. Below, we refer to each process (Section \ref{inj}, \ref{radipro}),
validity (Section \ref{validity}) and implication (Section \ref{maglim}) of the assumptions, and calculational method (Section \ref{method}).

\subsection{Injection of electrons}
\label{inj}
To begin with, we assume that electrons are injected constantly at the location of the compact object.
This assumption may be justified under the specific model, e.g., the pulsar scenario in which electrons are accelerated 
near the pulsar magnetosphere \citep{sie07,sie08a,sie08b}, 
because electron energy distribution is independent from orbital phase in such an injection scenario.
On the other hand, the assumption may not be suitable for some models, e.g., the microquasar scenario in which electrons are accelerated
at shocks in the jet \citep{bed06,bed07,kha08},
because electron energy distribution can vary with a change in the separation since the mass transport rate varies with
the separation and thereby the jet power changes. Thus, an injection depends on such models, but we assume the constant injection just 
because of simplicity. 

In this paper, we discuss only the case that injected electrons have energy distribution with a single power law,
i.e., $N_{e,\text{inj}}(E_e) \propto E_e^{-p}$.
In addition, the maximum energy and the minimum energy of electrons are assumed as
$\gamma _{e,\text{max}} = 10^8$ and $\gamma _{e,\text{min}} = 3\times 10^3$, respectively.

\subsection{IC and $\gamma \gamma$ absorption processes}
\label{radipro}
Figure \ref{noabs} shows the anisotropic and unabsorbed IC spectra from electrons with steady energy distribution which are injected constantly with power-law distribution, $p = 2.0$ and $p = 2.5$ in the thermal radiation field from the companion star at $k_BT = 3.3$ eV with the finite size.
IC fluxes increase and photon indices decrease as the scattering angle $\alpha$ (Figure \ref{modpic}) becomes larger both for 
$p = 2.0$ and $p = 2.5$.
It should be noted that the flux at photon energy $E_{\gamma} \sim \text{GeV}$ depends on $\alpha$ strongly, while that at 
$E_{\gamma} \sim 1\text{ TeV}$ less depends 
on $\alpha$. In addition, Figure \ref{noabs} shows that flux variation 
with orbital phase increases as the inclination angle $i$ increases, because $\alpha = 150^{\circ} \text{ and } 30^{\circ}$ correspond to 
the collision angle at superior conjunction (SUPC) and inferior conjunction (INFC), respectively, when $i = 60^{\circ}$ (see Figure \ref{modpic}). 
In the same manner, $\alpha = 120^{\circ} \text{ and } 60^{\circ}$ correspond to that at SUPC and INFC when $i = 30^{\circ}$, 
and $\alpha = 105^{\circ} \text{ and } 75^{\circ}$ correspond to that of $i = 15^{\circ}$ (see Figure \ref{modpic}). 
Moreover, the ratio of GeV flux to TeV flux is large when $p = 2.5$, while it is small when $p=2.0$

It is inevitable for a high energy photon to be absorbed in the stellar radiation field. 
Figure \ref{tau} shows the optical depth for $\gamma \gamma$ absorption. According to this figure, the optical depth increases as the angle 
$\alpha$ increases, and the photon energy at the maximum $\tau$ shifts from $\sim$ 1 TeV to $\sim$ 100 GeV. Furthermore, 
when $i \le 30^{\circ}$ ($60^{\circ} < \alpha < 120^{\circ}$), $\tau$ exceeds 
unity in any orbital phase. Thus, $\gamma \gamma$ absorption is very efficient in LS5039, so that electromagnetic cascade develops in
the case where the synchrotron energy loss of electrons can be ignored.

\subsection{Cooling timescale and effect of the stellar wind}
\label{validity}
The emission from the system LS5039 can be considered as a stationary one if a divided phase has a range of a tenth of one orbital period.
This is meaningful when electrons whose energy range is $3\times 10^3 < \gamma_e < 10^8$ cool completely on much shorter time scale than 
that phase range.
The ratio of IC cooling time for the electrons in the Thomson regime $t_{\text{IC,T}}$ to a tenth of an orbital period $T_{\text{orb}} / 10 \sim 
3.4\times 10^4$ s is
\begin{equation}
\frac{t_{\text{IC,T}}}{T_\text{orb}/10} \sim 4\times 10^{-4} \biggl( \frac{\gamma _e}{3\times 10^3} \biggr)^{-1}
\left( \frac{a}{a_{\text{peri}}} \right)^2.
\label{tict}
\end{equation}
On the other hand, high energy electrons cool in the Klein-Nishina regime, so that
\begin{equation}
\frac{t_{\text{IC,KN}}}{T_\text{orb}/10} \sim 10^{-2} \biggl( \frac{\gamma _e}{10^8} \biggr)^{0.7}
\left( \frac{a}{a_{\text{peri}}} \right)^2,
\label{tickn}
\end{equation}
where $t_{\text{IC,KN}} \sim 22\ (a/a_{\text{peri}})^2(E_e/1\text{TeV})^{0.7} \text{s}$ is the cooling time of electrons in the Klein-Nishina regime
\citep{aha06b}.
Thus, the electrons with $3\times 10^3 < \gamma_e < 10^8$ lose energy sufficiently fast in the divided phase, and therefore
it is a good approximation that the emission from the system is stationary.

It is necessary that advection of electrons by the stellar wind can be ignored in order to validate our assumption of negligible propagation of
electrons. The ratio of the advection length during IC energy loss to the scale length of the system is
\begin{equation}
\frac{v_{w}t_{\text{IC,KN}}}{L_{\text{sys}}} \sim 10^{-1} \left( \frac{v_w}{3 \times 10^8 \text{cm s}^{-1}} \right) 
\biggl( \frac{\gamma _e}{10^8} \biggr)^{0.7} 
\left( \frac{a}{a_{\text{peri}}} \right)^2 \left( \frac{L_{\text{sys}}}{0.1 \text{AU}} \right)^{-1},
\end{equation}
where $v_w$ is the wind velocity of a massive star and an electron with $\gamma_e = 10^8$ is thought, for it cools more rapidly than an electron with 
$\gamma_e = 3\times 10^3$ (equations (\ref{tict}) and (\ref{tickn})).
Thus, the assumption of trapped electrons is a good approximation of electrons with $3\times 10^3 < \gamma _e < 10^8$. 

\subsection{Constraint on magnetic field}
\label{maglim}
Magnetic field is limited by the assumptions about cooling and propagation of electrons and positrons.
The former gives an upper bound because if the magnetic field is larger than a certain value, synchrotron loss is a dominant process, and the latter gives a lower bound because if it is smaller than a certain value, the gyro radius of an electron with the highest energy is comparable to the scale length of the system.
In order that the assumption that electrons lose energy only by IC process is valid, cooling times of IC ($t_{\text{IC}}$) and 
synchrotron ($t_{\text{sy}}$) process must satisfy the condition, 
\begin{equation}
t_{\text{sy}} > t_{\text{IC}}. 
\label{cootim}
\end{equation}
It is sufficient to give this condition
at electron energy $\sim 30\text{ TeV}$, which corresponds to the energy of electrons radiating the photon with highest energy observed, 
and electrons with such high energy lose energy in the Klein-Nishina regime.
Thus, equation (\ref{cootim}) is reduced to,
\begin{equation}
B \lesssim 0.2 \left( \frac{E_e}{30\text{TeV}} \right)^{-0.85} \left( \frac{a}{a_{\text{peri}}} \right)^{-1} \text{G},
\label{magcoo}
\end{equation}
at periastron, where $E_e$ is the electron energy, $a$ is the separation, and $a_\text{peri}$ is $a$ at periastron.
At the same time, $B \lesssim 0.1$ G at apastron, for $a_{\text{apa}} \sim 2 a_{\text{peri}}$.

In order that the propagation of electrons can be ignored, the gyro radius of electrons must be smaller than the scale 
of the system, i.e.
$E_e / eB \lesssim L_{\text{sys}} \sim 0.1 \text{AU}$.
This leads to,
\begin{equation}
B \gtrsim 0.1 \left( \frac{\gamma_e}{10^8} \right) \text{G}.
\label{magprop}
\end{equation}
Therefore, under the condition of the uniform magnetic field, the equations (\ref{magcoo}) and (\ref{magprop}) give $B \sim 0.1$ G. 

\subsection{Method of calculation}
\label{method}
We calculate phase-divided spectra and light curves emerging from the system by IC and $\gamma \gamma$ absorption as well as the resultant
energy distribution of electrons.
We performed a numerical calculation of IC scattering and the subsequent cascade under the condition of the fixed separation 
by Monte Carlo method based on the stationary emission discussed above, thereby we determined the energy and the direction of the escaped photons. 

In the Monte Carlo procedure, physical quantities are determined by weighted random numbers in each process. First of all, the energy and the direction
of an injected electron are determined by random numbers. Second, the energy and the direction of a target photon in IC process in the observer
frame are determined by random numbers, which in turn determine the energy of the incident photon in the electron rest frame. The energy and 
the direction 
of the scattered photon in the electron rest frame is determined by physical quantities before the scattering and by one random number with the 
distribution of the differential cross section. These physical quantities after the scattering determine the energy of the scattered photon in 
the observer frame. We assume that the direction of this outgoing photon equals that of the injected electron because
injected electrons have extremely relativistic energy in our model. The location of annihilation of the photon, the energy and the direction of 
a partner photon in photon-pair annihilation process, and the direction of the $e^{\pm}$ pair in the center of $e^{\pm}$ mass frame are also
determined with random numbers. The energy and the direction of the $e^{\pm}$ pair in the center of $e^{\pm}$ mass frame determine their energy 
in the observer frame and they cause IC scattering again. We can calculate the radiation transfer taking into account the cascade process 
by iterating such procedure.

Synchrotron spectra are calculated by assuming the uniform magnetic field and using the resultant energy distribution of electrons and positrons. 
We can obtain their energy distribution by weighting the energy distribution
resulted by Monte Carlo method by their cooling time. Therefore, assuming the value of magnetic field, we can derive the synchrotron spectra.
Thus, we can calculate the spectra and the light curves of a circular-orbit approximation in keV, GeV, and TeV energy ranges, 
taking as parameters the radius of the circular orbit, namely, the separation $a$, inclination angle $i$, and the power-law index $p$ of the injected electrons. 

\section{Results}
\label{result}
\subsection{Inverse Compton spectra and light curves}
\label{sepa}
\begin{figure}
\begin{center}
\includegraphics[height=6cm,angle=270,clip]{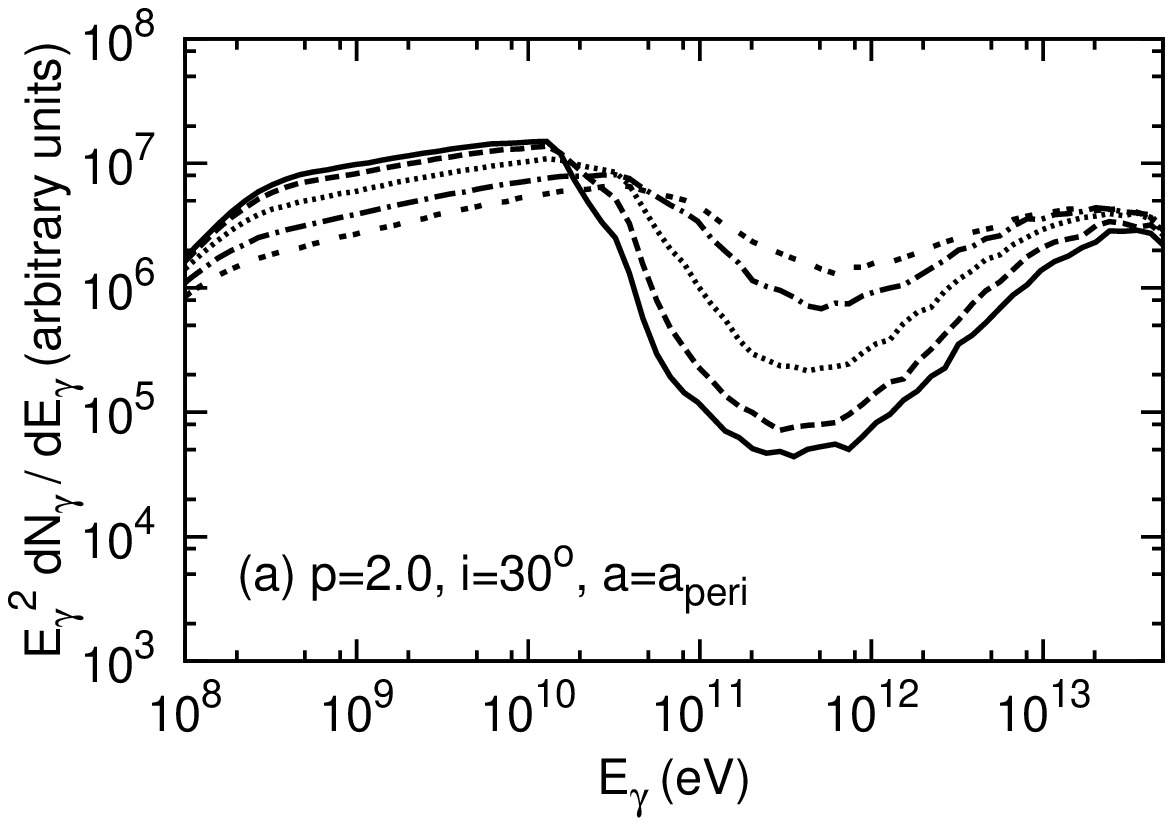}
\includegraphics[height=6cm,width=4cm,angle=270,clip]{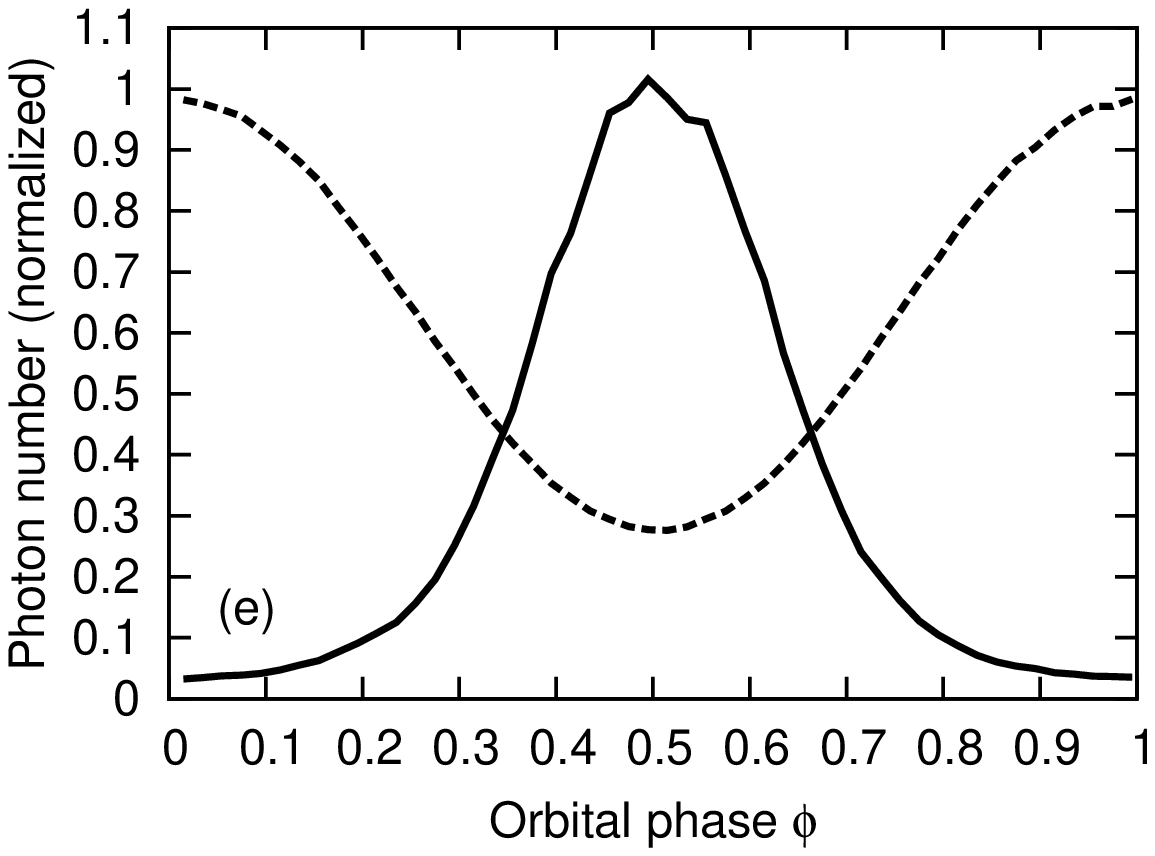}
\includegraphics[height=6cm,angle=270,clip]{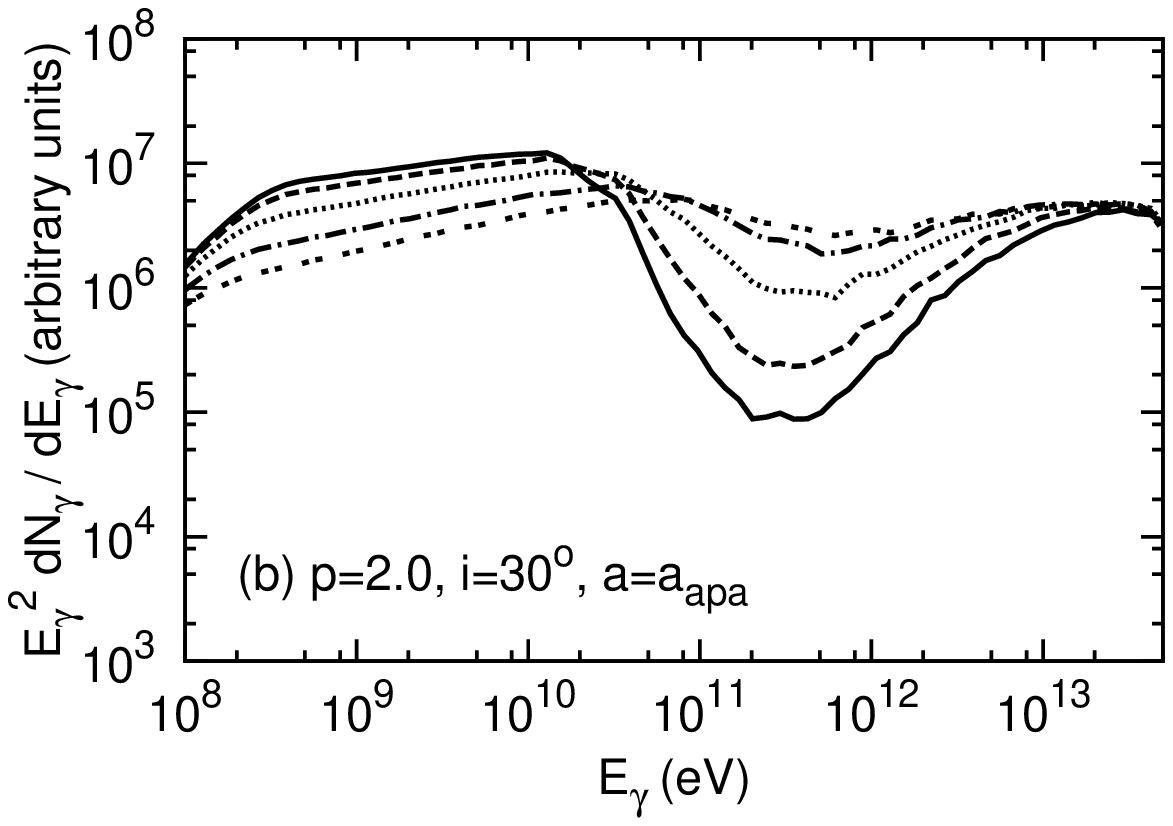}
\includegraphics[height=6cm,width=4cm,angle=270,clip]{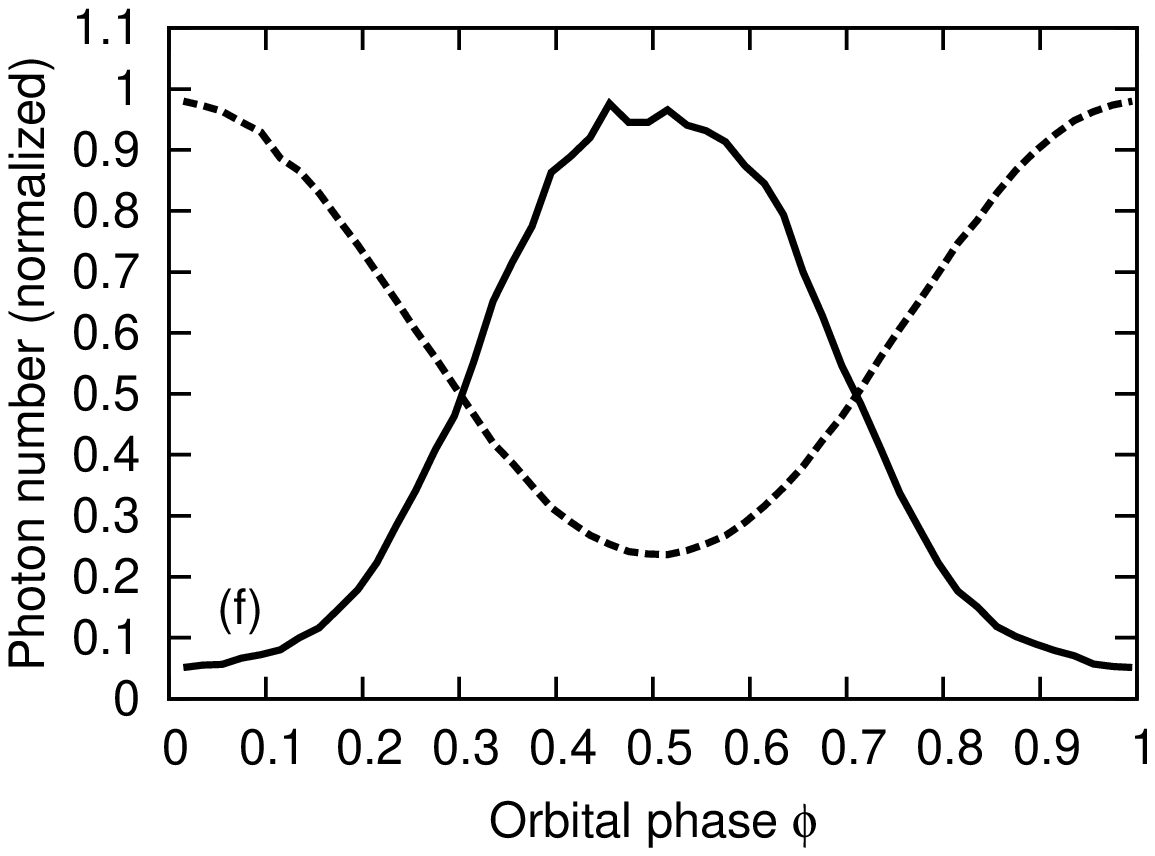}
\includegraphics[height=6cm,angle=270,clip]{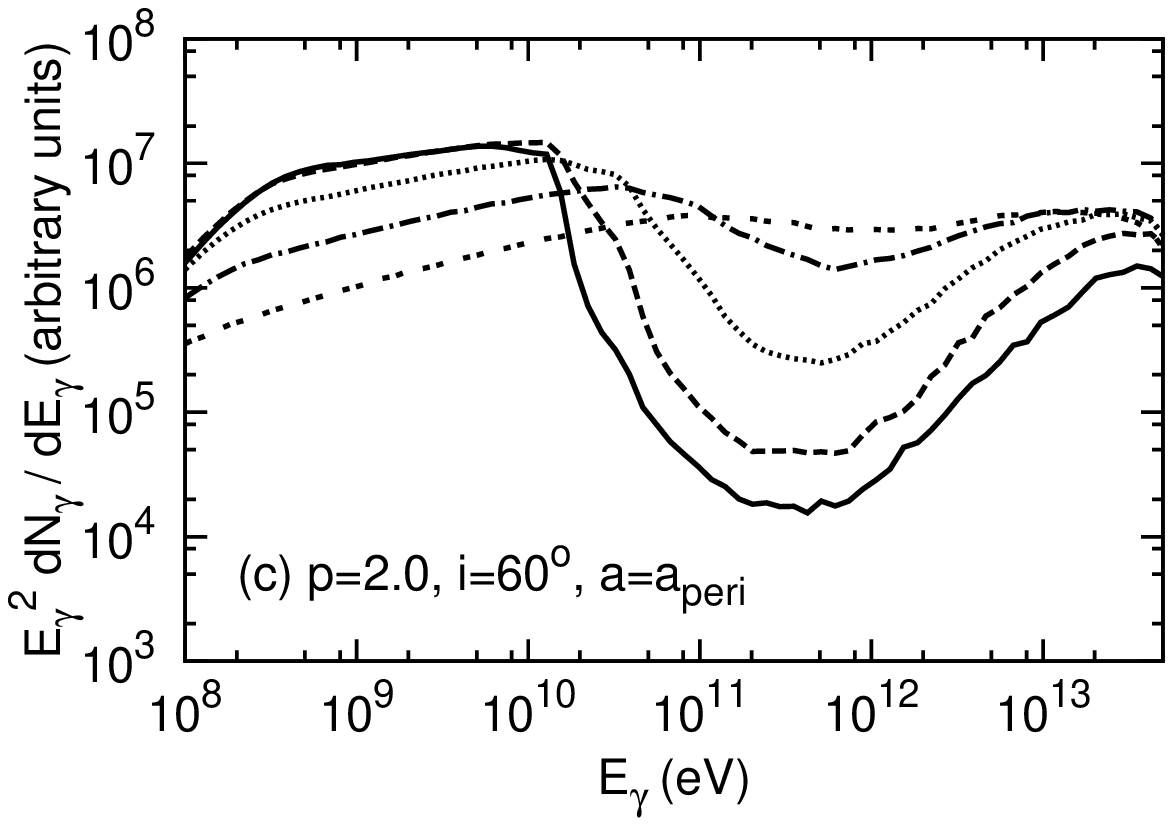}
\includegraphics[height=6cm,width=4cm,angle=270,clip]{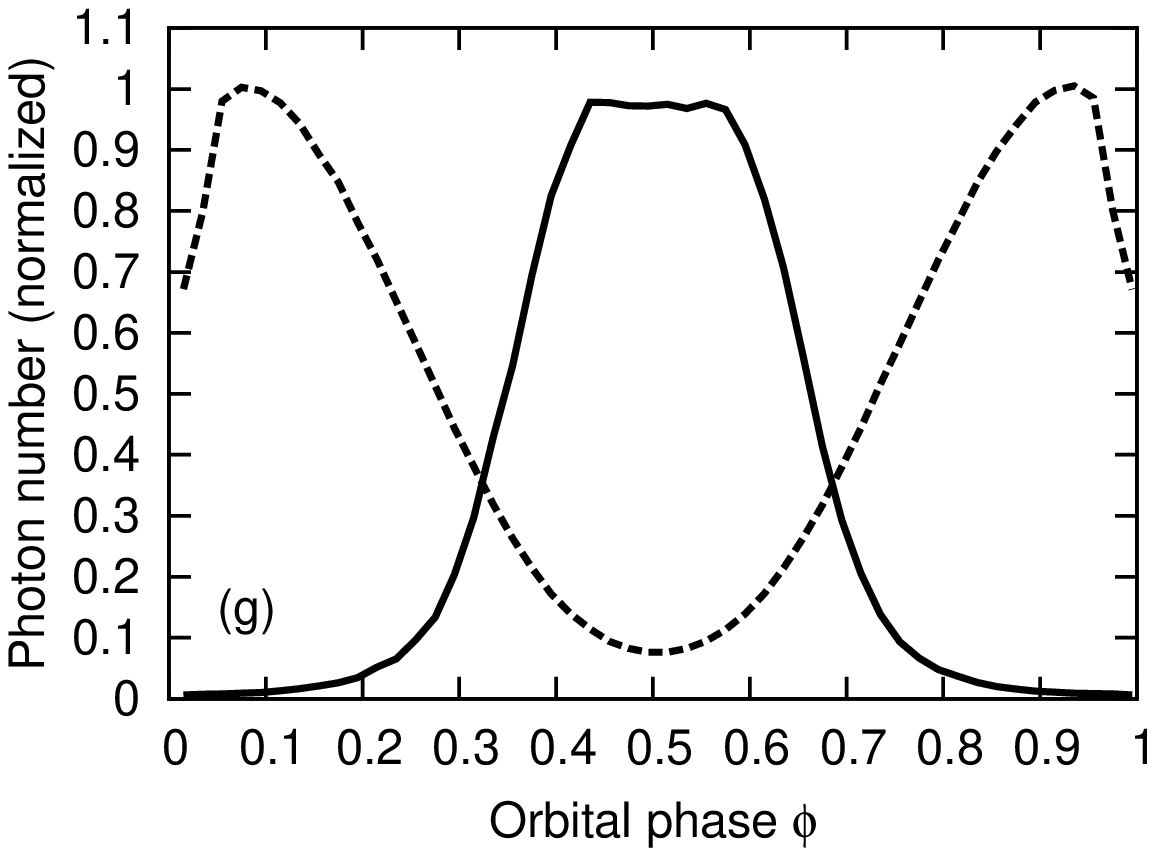}
\includegraphics[height=6cm,angle=270,clip]{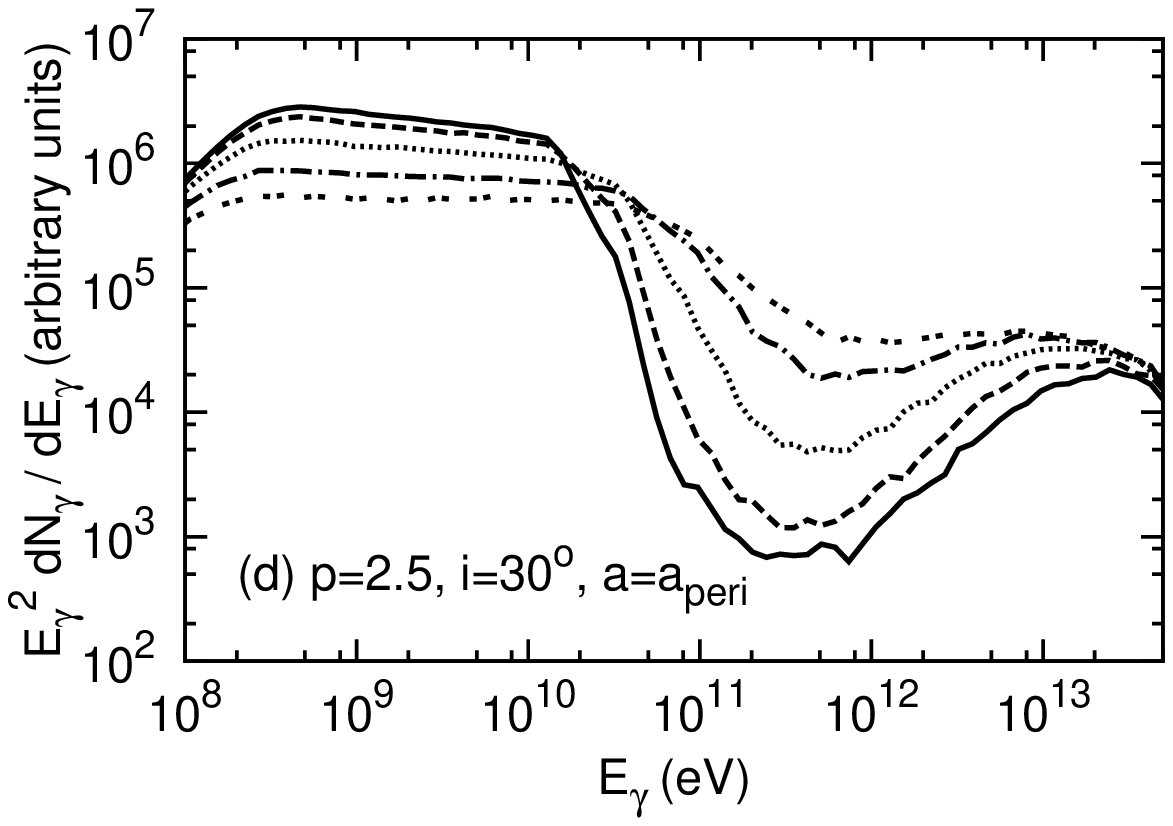}
\includegraphics[height=6cm,width=4cm,angle=270,clip]{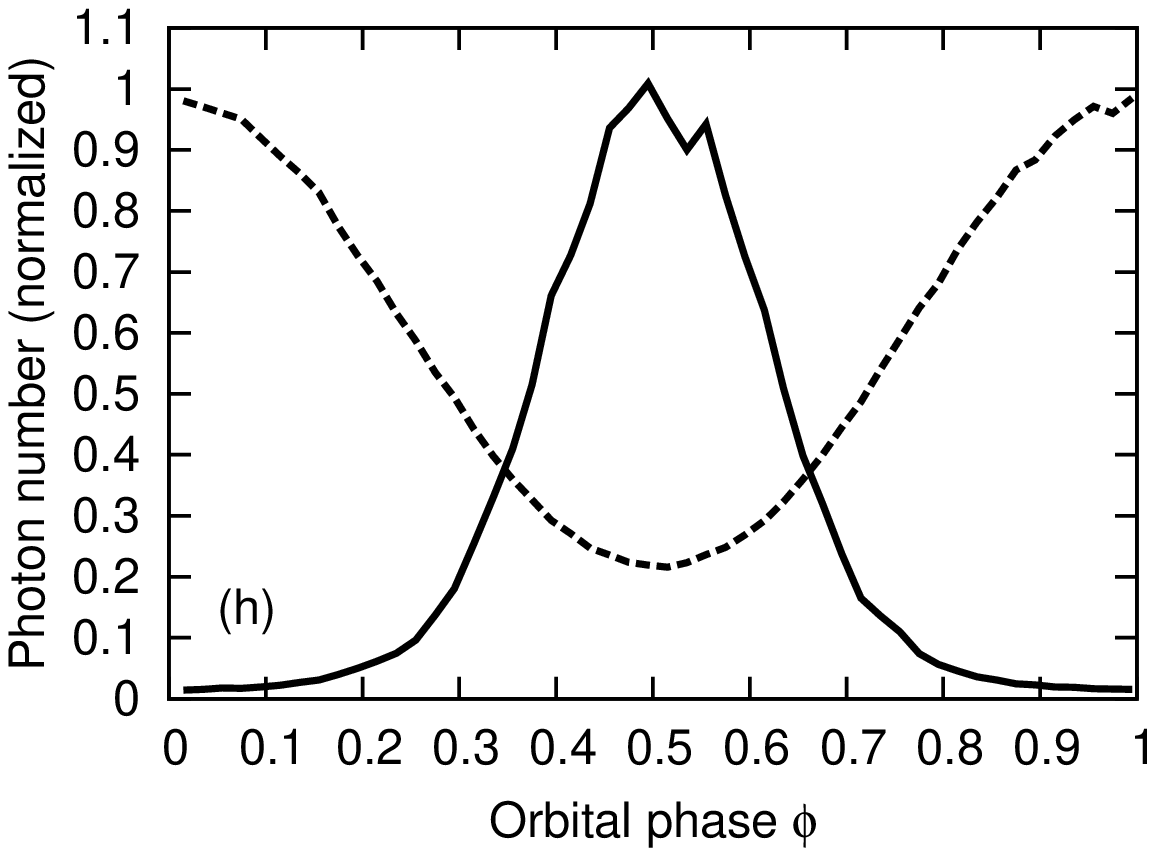}
\caption{Phase divided spectra (left panels; a, b, c, and d) and light curves (right panels; e, f, g, and h) for the circular orbit. 
In showing the spectra, orbital phases are divided into five intervals, $\phi = 0\text{-}0.1$ 
(solid line), 0.1-0.2 (dashed), 0.2-0.3 (dotted), 0.3-0.4 (dot-dashed), 
and 0.4-0.5 (double-dotted), where $\phi = 0$ represents SUPC and $\phi = 0.5$ represents 
INFC. The light curves of fluxes in the energy range 0.2-5 TeV (solid lines in the right panels) and 0.1-10 GeV 
(dashed lines in the right panels) are shown, and each light curve is normalized so that its maximum is unity. The calculations are performed for
$p = 2.0, i = 30^{\circ}, a = a_{\text{peri}}$(a and e); 
$p = 2.0, i = 30^{\circ}, a = a_{\text{apa}}$(b and f); 
$p = 2.0, i = 60^{\circ}, a = a_{\text{peri}}$(c and g); 
$p = 2.5, i = 30^{\circ}, a = a_{\text{peri}}$(d and h).}
\label{spe}
\end{center}
\end{figure}

The results of the Monte Carlo calculation are shown in Figure \ref{spe}, where we assume that the binary orbit is a circular one. 
The phase-divided spectra by IC cascade process are shown in the left panels in Figure \ref{spe} and the light curves of fluxes in TeV band
and GeV band are shown in the right panels in Figure \ref{spe}. The spectra and light curves shown in Figures \ref{spe}a and e are calculated for
$p = 2.0, i = 30^{\circ}, a = a_{\text{peri}}$, and that in Figures \ref{spe}b and f, Figures \ref{spe}c and g, and Figures \ref{spe}d and h 
are calculated, changing the separation to $a=a_{\text{apa}}$, the inclination angle to $i=60^{\circ}$, 
and the power-law index of distribution of injected electrons to $p=2.5$, respectively.

\begin{figure}
\begin{center}
\includegraphics[height=8cm,angle=270,clip]{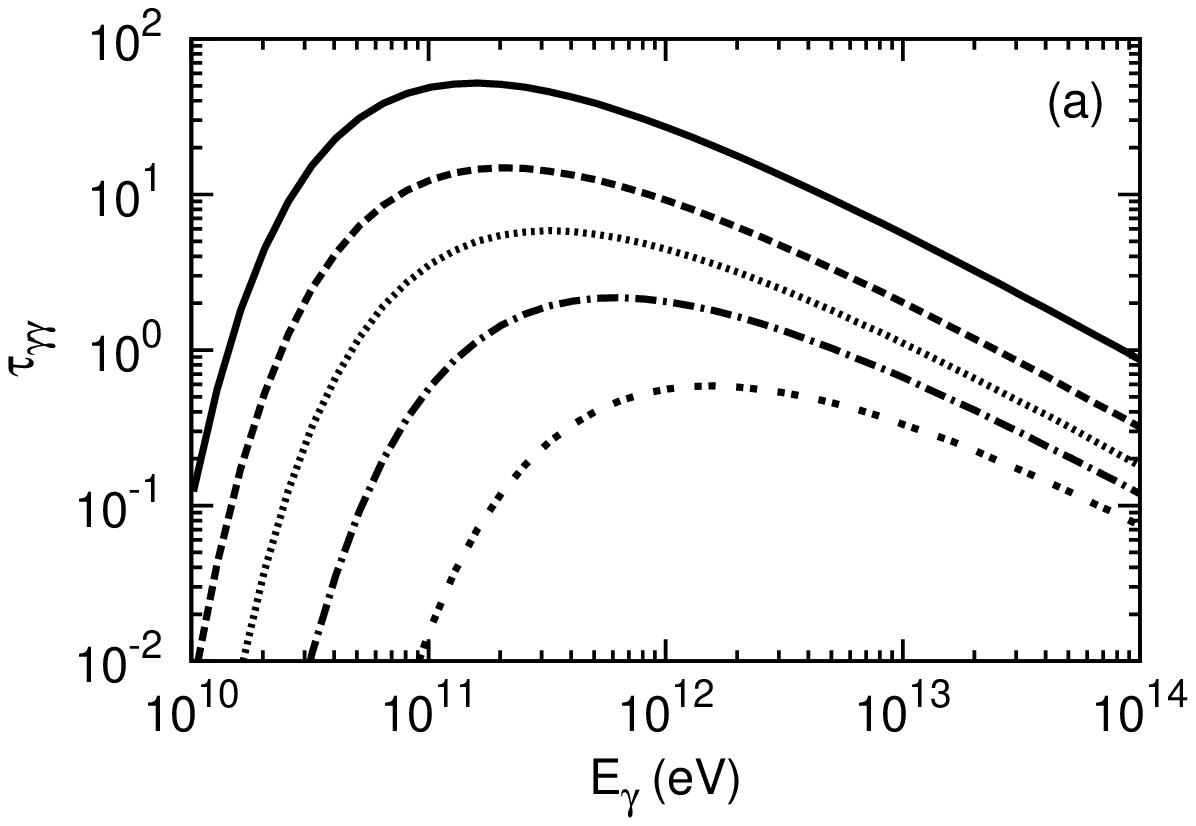}
\includegraphics[height=8cm,angle=270,clip]{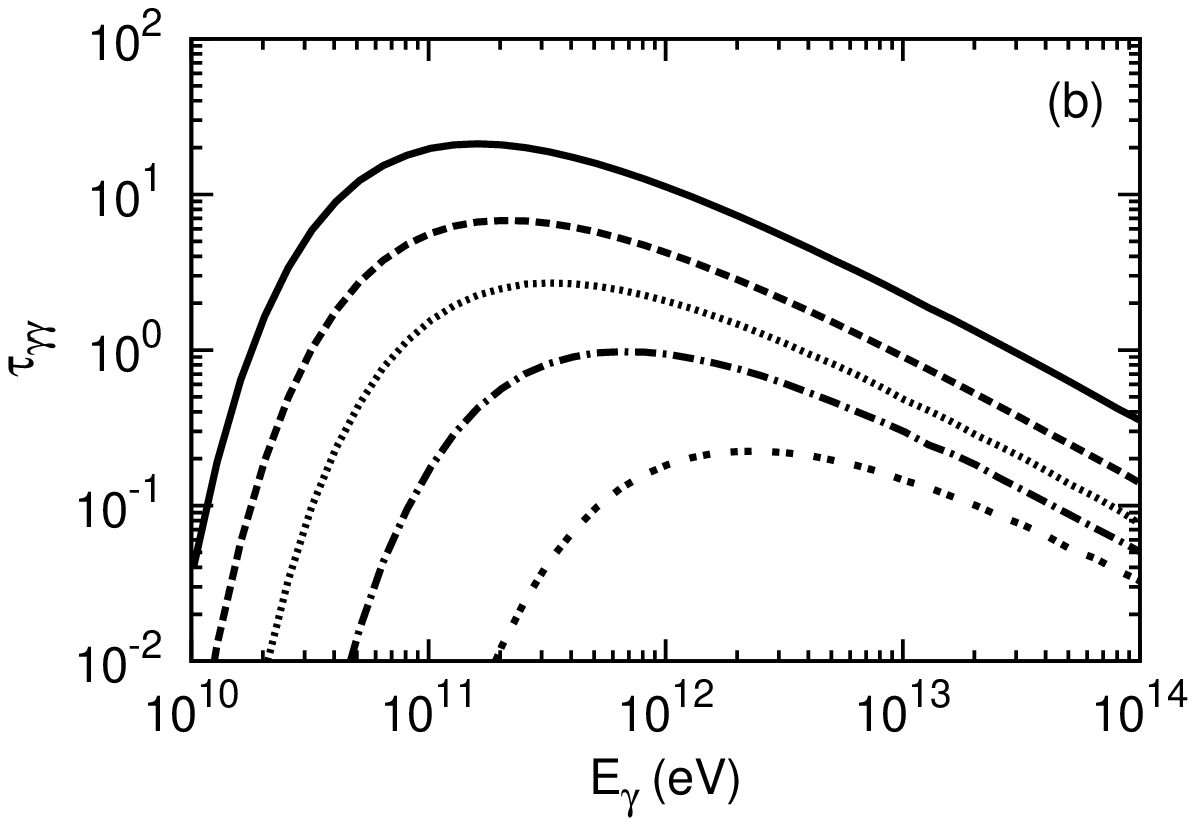}
\caption{Optical depth for high energy photons in the stellar radiation field at periastron (a) and apastron (b).
The origin of the photon path is the location of the compact object, and the calculation is performed in the case that the angle $\alpha$
is $150^{\circ}$ (solid line), $120^{\circ}$ (dashed), $90^{\circ}$ (dotted), 
$60^{\circ}$ (dot-dashed), and $30^{\circ}$ (double-dotted).}
\label{tau}
\end{center}
\end{figure}

First of all, we discuss the results shown in Figures \ref{spe}a and e.
There are some remarkable points in the spectra of GeV and TeV in the case of 
$p = 2.0, i = 30^{\circ}, a = a_{\text{peri}}$ (Figure \ref{spe}a). 
In TeV energy range, the photon indices are smaller than 2 at all divided phases
since the optical depth decreases monotonically with the photon energy (Figure \ref{tau}) and since 
the spectrum without $\gamma \gamma$ absorption has a flat feature in TeV energy range (Figure \ref{noabs}a).
The photon energy at which the flux in the TeV range becomes minimum is different from that at which the optical depth becomes maximum (e.g. the solid line in Figure \ref{spe}a and the dashed line in Figure \ref{tau}a), though we expect naively that these two kinds of energy are equal. The gap is due to the cascade process, which moves the photon energy to lower energy band, so that the photon energy of the minimum flux shifts to higher energy. 
On the other hand, in GeV energy band IC photons are not subject to the $\gamma 
\gamma$ absorption in the stellar radiation field (Figure \ref{tau}a), but the photon indices of all phases slightly increase compared with the 
IC spectra without absorption (Figure \ref{noabs}) because of cascade process.
Another noticeable fact is that the flux at energy $\gtrsim$ 10 GeV decreases in each orbital phase, since the optical depth rapidly increases at $\sim$ 100 GeV.
As for light curves under the same condition (Figure \ref{spe}e), GeV flux varies inversely with TeV flux 
because of the anisotropic IC radiation in GeV and because of the efficient absorption in TeV 
as can be expected from the spectra.
In addition, the amplitude of flux variation in GeV range is approximately a factor 3, 
while that in TeV greatly exceeds one order of magnitude, consistent with \citet{bed06}.

For $p = 2.0, i = 30^{\circ}, a = a_{\text{apa}}$ (Figures \ref{spe}b and f), 
the spectra and the light curves in energy range of 0.1-10 GeV are almost the same as in the case 
of $a = a_{\text{peri}}$. 
When the compact object is located far from the companion star, the stellar radiation field at the compact object becomes thinner, so that IC cooling time of electrons becomes longer because of less reaction rate of IC scattering. Thus, if the injection rate 
of electrons is constant, the electron number density at the same energy becomes larger.
On the other hand, the GeV flux is proportional to the number density of electrons because in 0.1-10 GeV energy range, 
photons are not absorbed in the thermal radiation field.
Therefore, the flux, which is equivalent to the product of the power by IC process 
per one electron and the electron number, is as large as that at periastron in spite of 
thinner field. However, the GeV flux is actually subject to the cascade process as stated above. This means that the amplitude of flux variation
in GeV band can change with the amount of energy moving to lower energy band by cascade process, which changes with the separation because the amount of energy of absorbed photons changes with the separation. Nevertheless,
the amplitude of GeV flux in the case $a=a_{\text{apa}}$ (dashed line in Figure \ref{spe}f) is almost the same as 
in the case $a=a_{\text{peri}}$ (dashed line in Figure \ref{spe}e). This implies less impact of cascade on variation of GeV flux.
Thus, flux variation in that energy range is determined almost exclusively by the anisotropic IC radiation.
In contrast, TeV flux becomes larger than the case of $a=a_{\text{peri}}$, because the optical depth 
becomes smaller by thinner radiation field, but there is almost no change in the amplitude of flux variation.

Observing the system with the inclination angle $i = 60^{\circ}$ (Figures \ref{spe}c and g), one notices that 
the amplitudes of flux variations in both TeV and GeV range become larger than the case 
of $i = 30^{\circ}$. The increasing amplitudes of flux in GeV range and TeV range are explained 
by anisotropic IC process (Figure \ref{noabs}a) and $\gamma \gamma$ absorption (Figure \ref{tau}) which 
are dependent on the angle $\alpha $ (Figure \ref{modpic}). 
When the inclination angle becomes larger, the photon path from the compact object passes through the thicker radiation field at SUPC phase,
and the thinner field at INFC, so that the amplitude of TeV flux becomes larger. The flux by anisotropic IC process varies more largely
with the orbital phase, so that the amplitude of GeV flux also becomes larger.
The decrease of GeV flux at SUPC phase (dashed line in Figure \ref{spe}g) is due to the collision of IC photon with the stellar surface. In addition,
the light curve of TeV flux around INFC phase has a flat feature (solid line in Figure \ref{spe}g) because the optical depth is smaller than unity
in that phase.

We also studied the case $p = 2.5$ (Figures \ref{spe}d and h), in which TeV to GeV flux ratio decreases
by one order of magnitude, compared to the case $p = 2.0$, while the amplitude of flux variation do not change very much. 

\subsection{Synchrotron spectra}
\label{ressyn}
\begin{figure}
\begin{center}
\includegraphics[height=8cm,angle=270,clip]{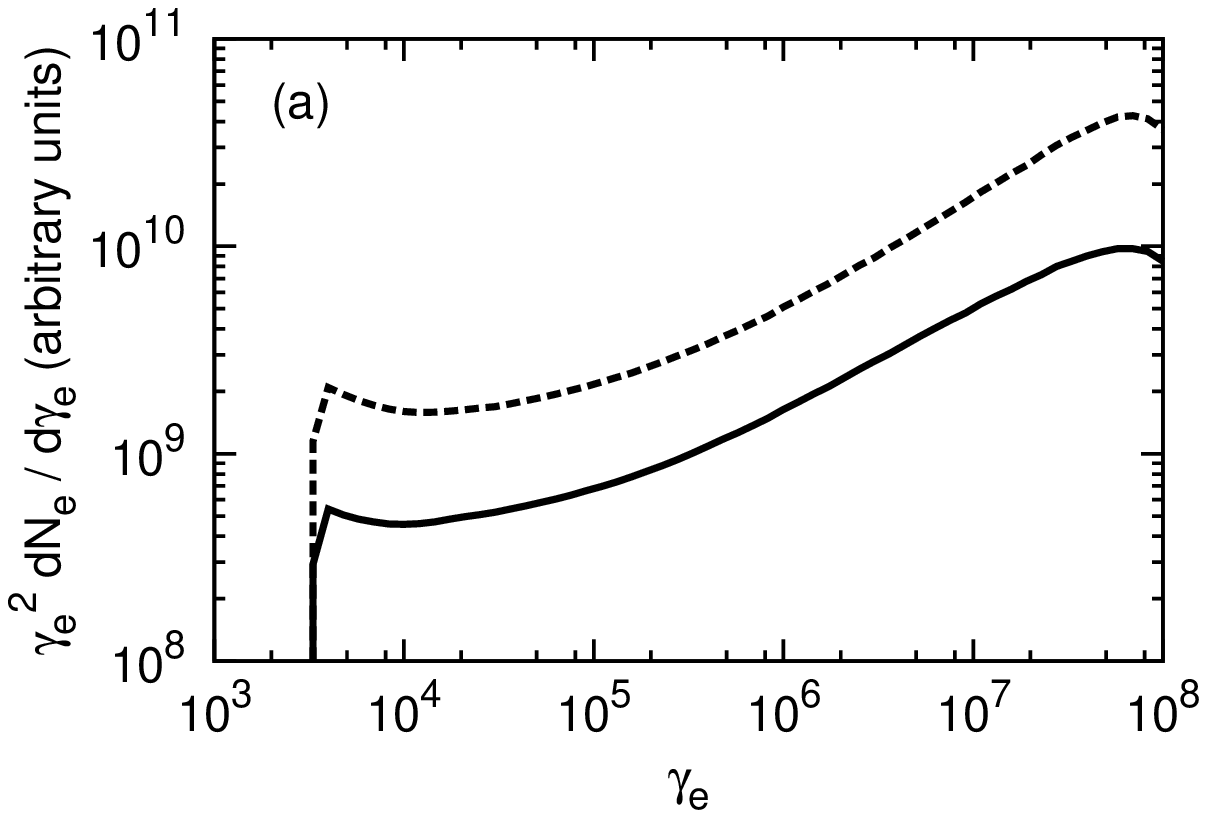}
\includegraphics[height=8cm,angle=270,clip]{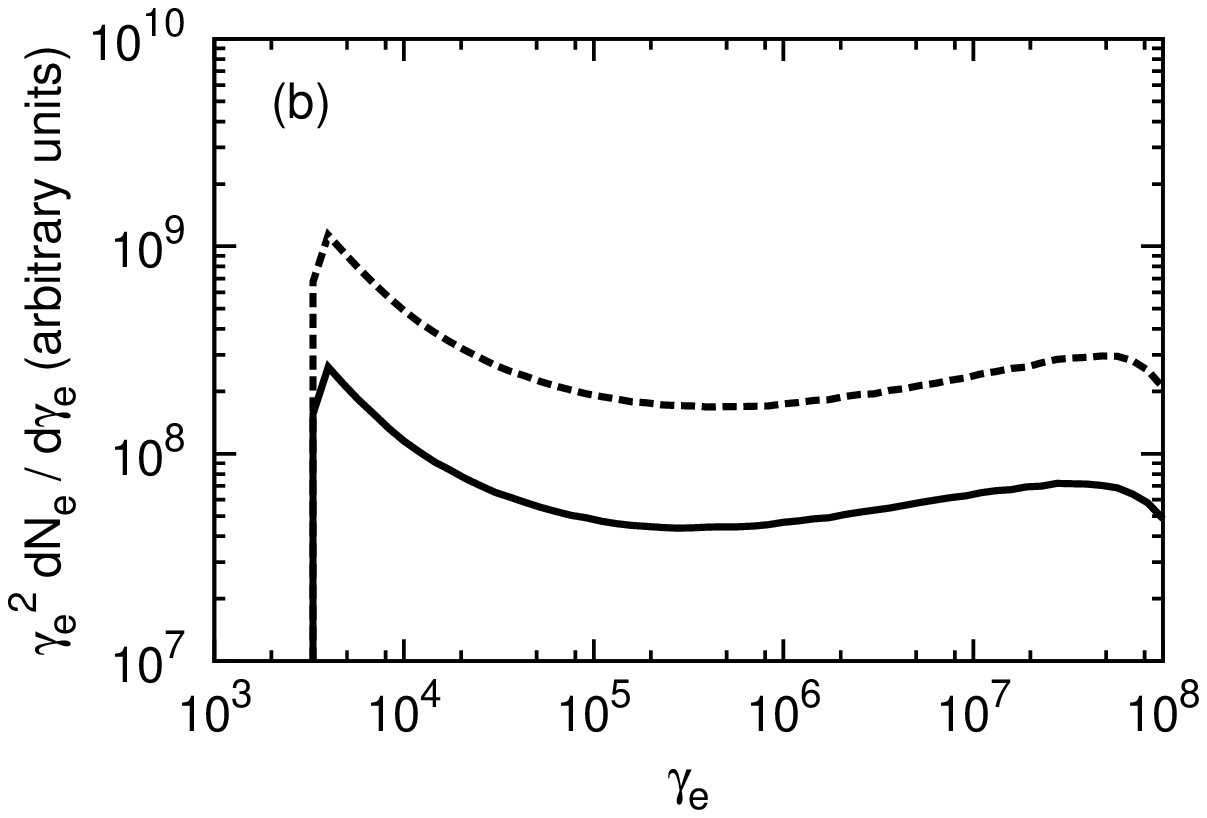}
\includegraphics[height=8cm,angle=270,clip]{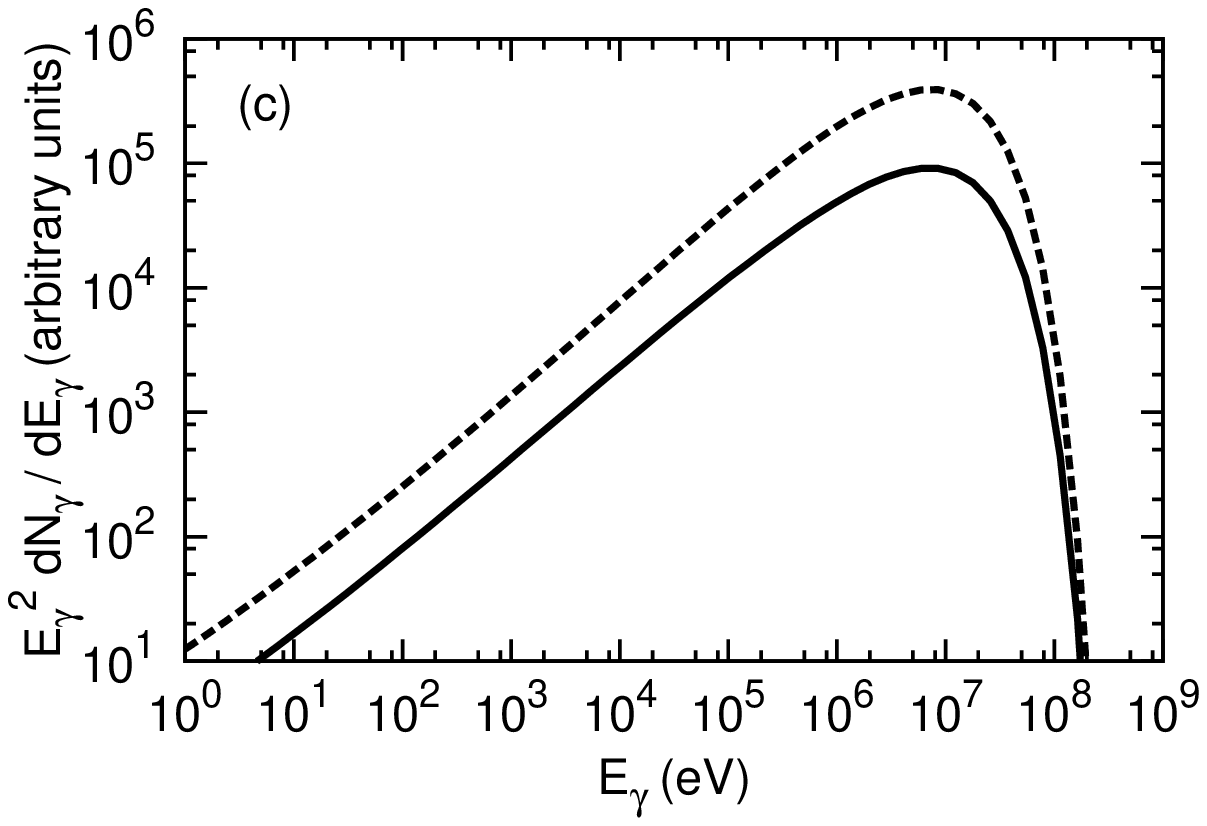}
\includegraphics[height=8cm,angle=270,clip]{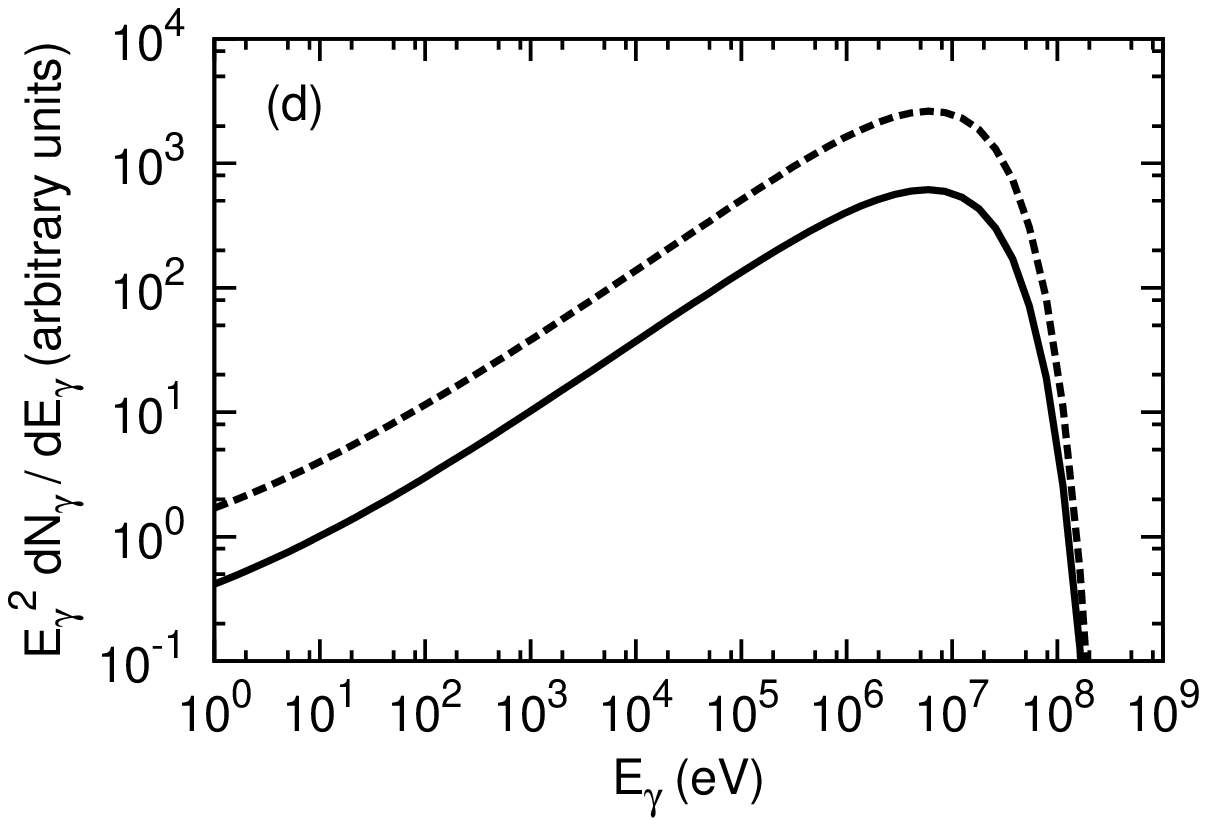}
\caption{Electron energy distributions in steady states (a and b) and synchrotron spectra (c and d). 
The electrons are injected with $p = 2.0$ (a and c) and $p = 2.5$ (b and d). 
The calculations are performed in the case $a = a_{\text{peri}}$ (solid line) and $a = a_{\text{apa}}$ (dashed).}
\label{syn}
\end{center}
\end{figure}

The electron distributions and the synchrotron spectra are shown in Figure \ref{syn}. 
The electron number density of the case $a = a_{\text{apa}}$ (dashed line in Figures \ref{syn}a and b) is 3 or 4 times larger 
than that of the case $a = a_{\text{peri}}$ (solid line in Figures \ref{syn}a and b) at any energy range, 
because the IC cooling time is inversely proportional to the energy density of the stellar radiation field (note that we consider the case 
the injection rate is constant).
As a result, the flux by synchrotron radiation in the case $a = a_{\text{apa}}$ is about 4 times larger (Figures \ref{syn}c and d). 

Another important feature is the power-law index of electron distribution and synchrotron spectrum. When $p = 2.0$, the index of electron distribution 
at $\gamma _e \sim 10^6$, an electron with which radiates photons with keV energy, is $\sim 1.5$ because of the Klein-Nishina effect and the contribution of 
secondary electrons, so that the photon index at keV is $\sim 1.3$. On the other hand, in the case $p = 2.5$, the relevant spectrum is almost 
flat, accordingly, the photon index is $\sim 1.5$.

\subsection{Comparison with observations}
\label{comp}
\begin{figure}
\centering
\includegraphics[height=13cm,angle=270,clip]{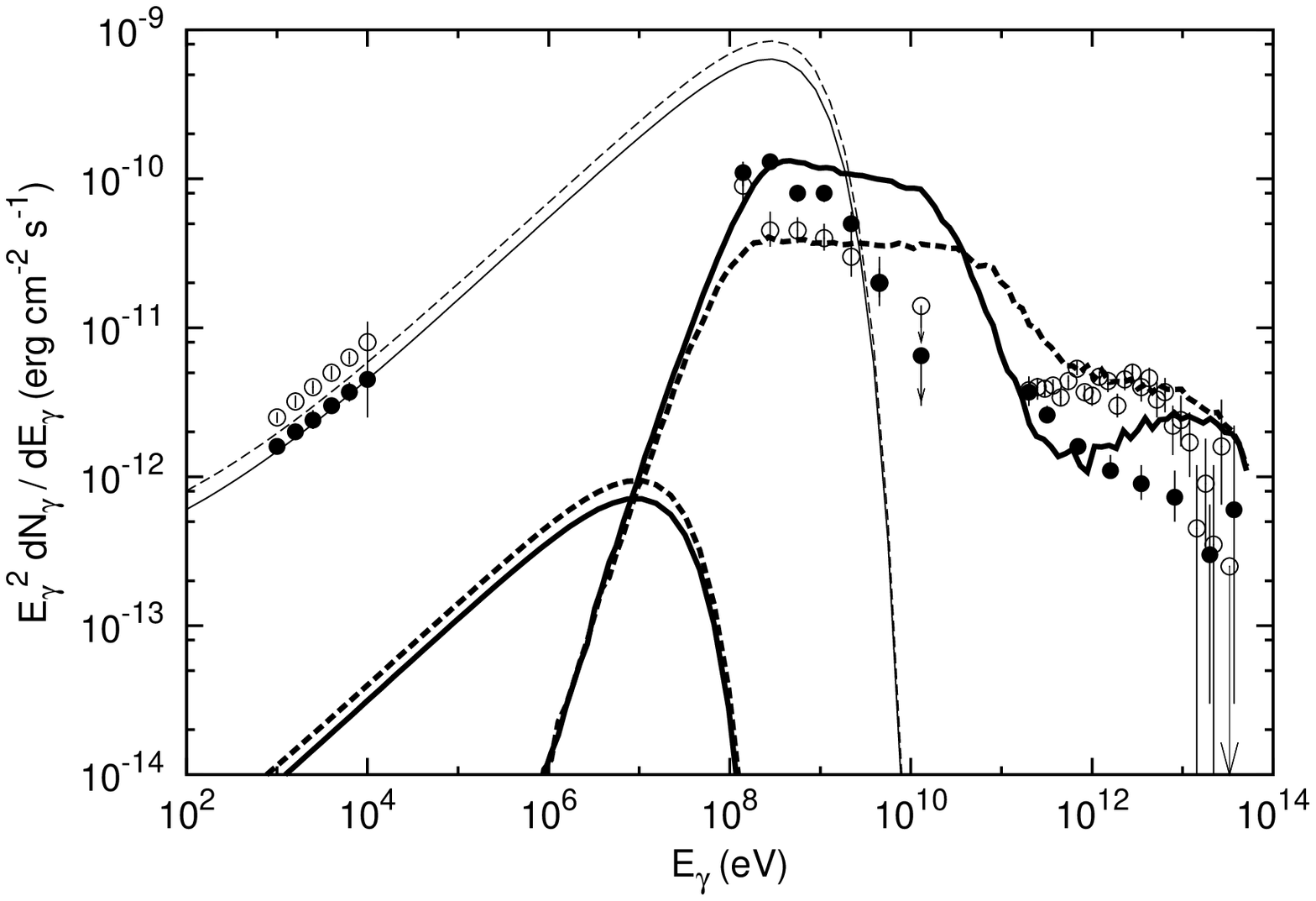}
\caption{Comparison of numerical results with the observational data obtained with Suzaku \citep[XIS,][]{tak09},
Fermi \citep{abd09}, and HESS \citep{aha06a}. The spectra averaged in the phases around SUPC 
(numerical: solid line, observed: filled circle) and around INFC (numerical: dashed, observed: open circle) are shown, where the orbital phase 
is divided into 0-0.45, 0.9-1.0 (around SUPC) and 0.45-0.9 (around INFC). The condition of the numerical 
calculation is $p = 2.5, i = 30^{\circ}, B = 0.1\text{ G}$ (thick line). We also show synchrotron spectra 
in the case of $B=3$ G (thin line) as a guide, assuming the power-law distribution of electrons shown in Figure \ref{syn}b.}
\label{cmprspe}
\end{figure}

\begin{figure}
\centering
\includegraphics[height=10cm,angle=270,clip]{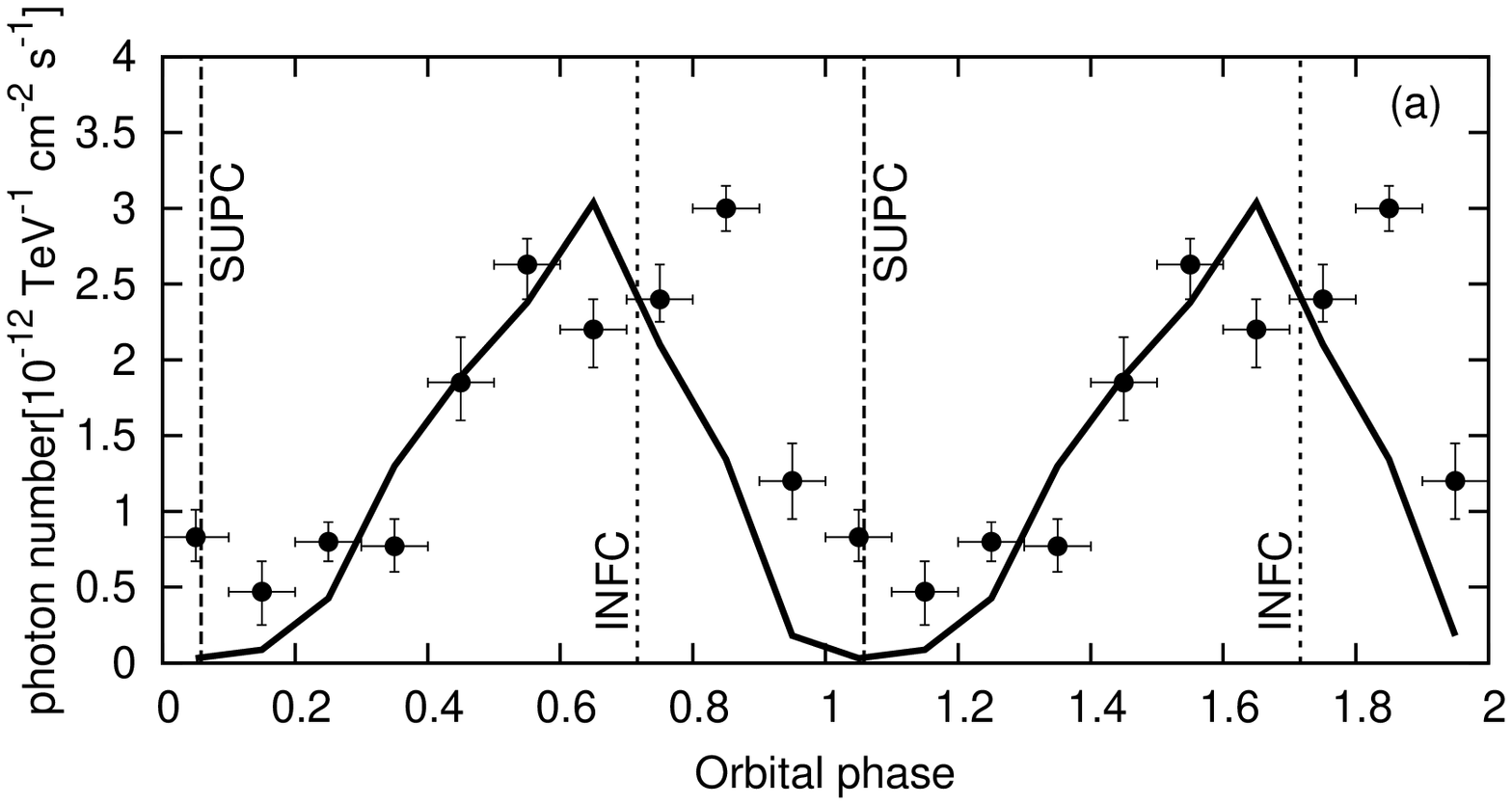}
\includegraphics[height=10cm,angle=270,clip]{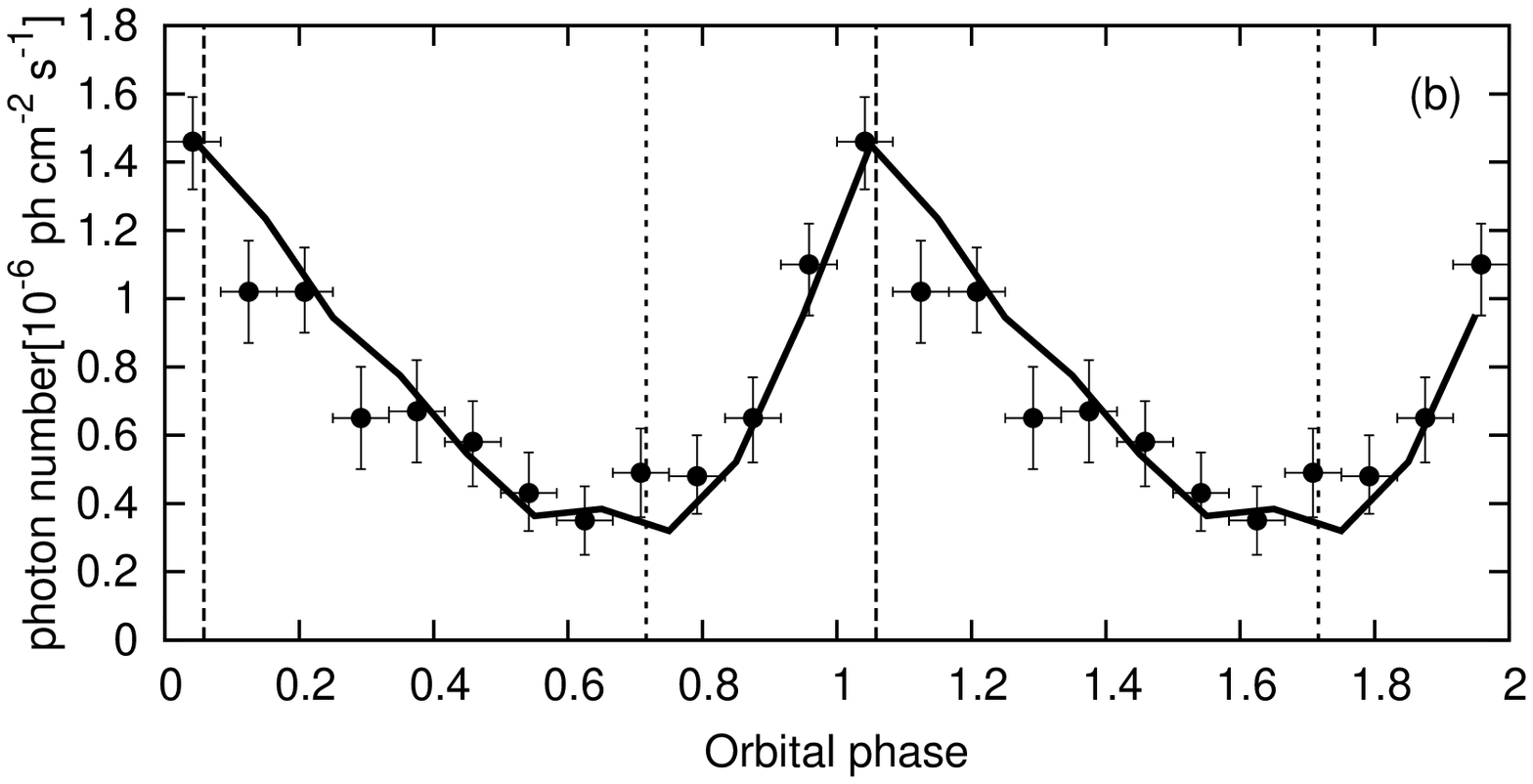}
\includegraphics[height=10cm,angle=270,clip]{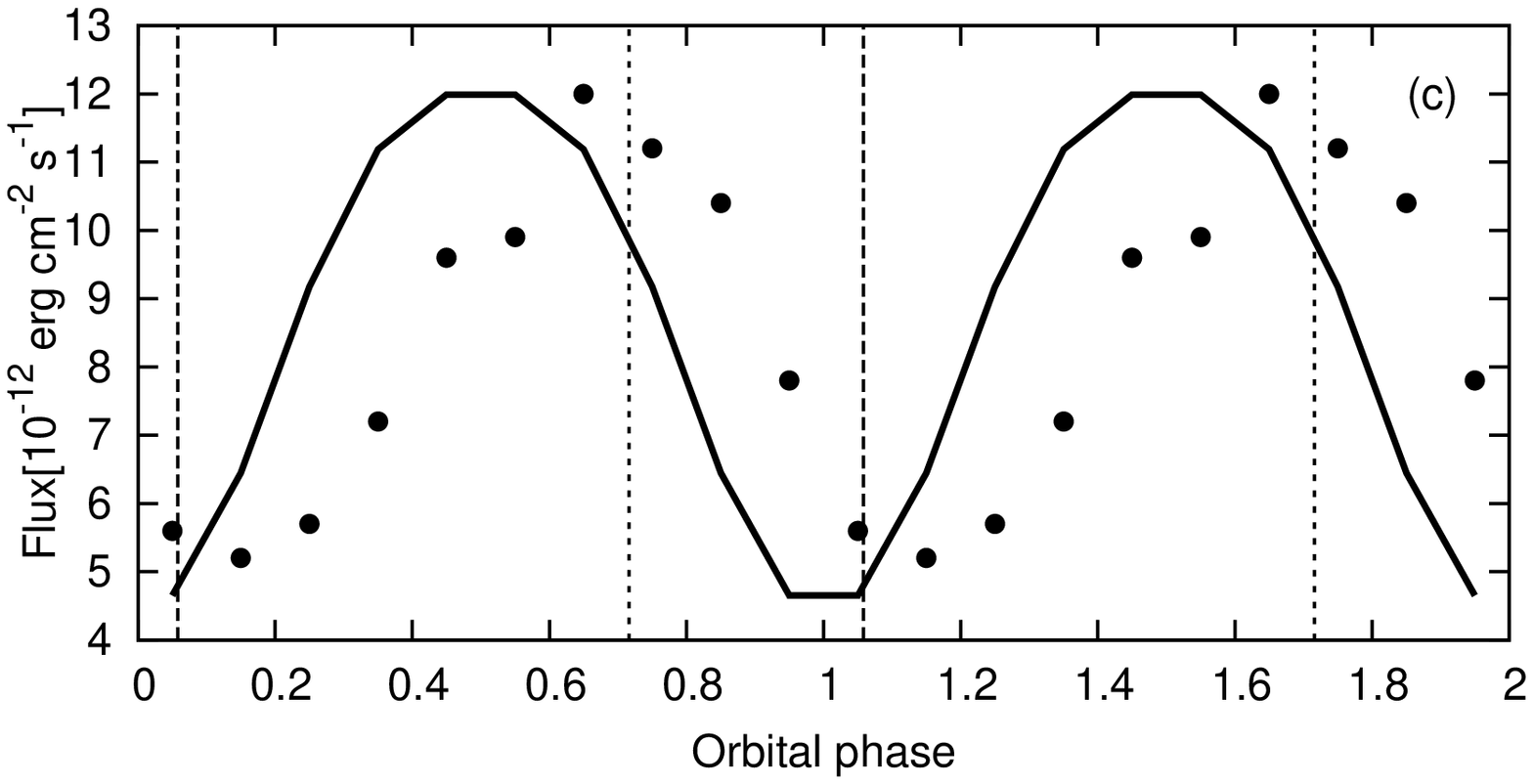}
\caption{Light curves of numerical results (solid line) and the observational data (circle with error bar). 
That of HESS energy band, 0.2-5 TeV \citep[a,][]{aha06a}, Fermi, 0.1-10 GeV \citep[b,][]{abd09}, 
and Suzaku, 1-10 keV \citep[c,][]{tak09} are shown.
Numerical result is normalized so that the maximum value matches the observed one. $\phi =0$ and $\phi=0.5$ represent a periastron phase and 
an apastron phase, respectively. SUPC (dashed line) and INFC (dotted line) phases are shown in each panel.}
\label{cmprlc}
\end{figure}

Taking into account the eccentricity of the binary system, we have also calculated spectra averaged around SUPC and INFC (Figure \ref{cmprspe}) 
and light curves (Figure \ref{cmprlc}) to compare with the observational data.
In calculating the spectra and the light curves, we divided the orbital phase into 10 equal parts, and gave the separation in each part
the average value of both ends. Thus, this calculation is made for five different separations.
The parameters are selected from the previous section so that the ratio of TeV flux to GeV one and flux variation in GeV range approximately match
the observational data. The spectra of $p=2.5$ better fit the observational data in terms of TeV to GeV flux ratio than $p=2.0$, and 
that of $i=30^{\circ}$ fit better in terms of the flux variation at the photon energy $\sim$ 1 GeV than $i=60^{\circ}$.

It is seen that the observed spectra can be reproduced reasonably well. That is, it matches the observation that the flux in INFC
phase is larger than that in SUPC as for X-ray and TeV energy ranges, and conversely INFC flux is smaller in GeV range. Moreover, serendipitously the 
photon index in X-ray range is reproduced. However, the spectra have several issues which do not reproduce the observations in detail. 
First of all, the model underpredicts the flux at the Suzaku energy range by two orders of magnitude for $B=0.1$ G, which is the most critical.
Second, it overpredicts the flux in 1-10 GeV and shows a break at higher energy than the Fermi data. Finally, for the spectra around INFC the model overpredicts the flux in 0.2-1 TeV while it matches the observation in the energy band $\gtrsim$ 1 TeV, and for the spectra around SUPC the model overpredicts the flux more than 1 TeV while it matches in 0.2-1 TeV.
The issue of keV range may be solved by increasing the magnetic field up to 3 G (see thin curves in Figure \ref{cmprspe}), 
which is consistent with the result in \citet{tak09}.
However, if so, the synchrotron cooling becomes the dominant process
for electrons with $\gamma _e \gtrsim 2 \times 10^6$, which radiates IC photons with $E_{\gamma} \gtrsim 1 \text{ TeV}$ and synchrotron photons 
with $E_{\gamma} \gtrsim 10^2 \text{ keV}$,
so that fluxes at energy $E_{\gamma} \gtrsim$ 1 TeV decrease and synchrotron spectra have a break at $E_{\gamma} \sim 10^2$ keV. 
This modification to the synchrotron spectra would resolve the issue that the numerical data of $B=3$ G overpredict the observational data 
in the energy range of 0.1-1 GeV. At the same time, these facts mean that the
flux of 0.1-10 GeV and 1-10 keV energy ranges would not be subject to synchrotron cooling, even if we set $B = 3$ G,
because it is the electrons with energy $E_e \gtrsim 1\text{ TeV}$ 
that are affected by synchrotron cooling, and because the electrons which radiate synchrotron photons with energy 1-10 keV have energy 
$E_e \lesssim 1\text{ TeV}$. Moreover, electrons which radiate 0.1-10 GeV photons by IC process has energy $E_e \lesssim 100 \text{GeV}$. 
On the other hand, when $B=3$ G it is hard for electrons to be accelerated up to several tens of TeV \citep[e.g.][]{kha08,tak09}.
The overprediction of the flux in 1-10 GeV range may be relevant to higher energy photons than 
the stellar radiation field. If stellar photons with energies of 10-100 eV exist, photons with energies of 1-100 GeV are subject to
$\gamma \gamma$ absorption, so that the discrepancy of the spectra in that energy band may be resolved.

Light curves also approximately match the observational data, that is, TeV-GeV anticorrelation and TeV-X correlation are reproduced (Figure \ref{cmprlc}).
The flux for $E_{\gamma} <$ 10 GeV is determined only by the angle $\alpha$ (Section \ref{sepa}), so that GeV flux correlates with 
$\alpha$. On the other hand, anisotropic IC spectra more than 100 GeV also correlate with $\alpha$, but the effect that it gives on cascade 
spectra is minor. Instead, $\gamma \gamma$ absorption is effective in that energy range (Figure \ref{tau}), 
so that TeV flux anticorrelates with $\alpha$. Thus, 
GeV and TeV fluxes anticorrelate. Moreover, the synchrotron flux correlates with the binary separation (Section \ref{ressyn}), and 
the binary separation in LS5039 is short around SUPC and long around INFC. Therefore, TeV and X-ray fluxes appear to correlate.

Looking closely at light curves, one finds that that in HESS band does not match the observation 
especially around SUPC (Figure \ref{cmprlc}a), while
that in Fermi band is well reproduced (Figure \ref{cmprlc}b).
That of X-ray appears to match the observational data in terms of variation, but there is an phase difference by $\Delta \phi \sim 0.15$ (Figure \ref{cmprlc}c). 
This is somewhat a puzzle and this problem is related to the variation amplitude between SUPC and INFC at the Suzaku energy band.

\subsection{Mass range of the compact object}
Using the results of flux variation in GeV range, we can put limitation on the mass of the compact object.

To begin with, the amplitude of flux variation is determined by the inclination angle. 
GeV flux, which is not subject to $\gamma \gamma$ absorption, is decided exclusively by the angle $\alpha$
under the condition of the constant injection, because it is independent of the separation (Section \ref{sepa}). 
Moreover, the inclination angle determines the amplitude (Section \ref{model}). Therefore, determining the 
inclination angle, we can give the decisive value to the amplitude.
In the case of LS5039, the inclination angle is $i = 15^{\circ} \text{-} 30^{\circ}$, 
since the observational data shows that the amplitude is a factor 3-6 (Figures \ref{noabs} and \ref{cmprlc}).

The limitation of the inclination angle is equivalent to a limitation of the mass of the compact object. Determining the value of mass function 
by observations, we can obtain the relation between the inclination angle, the mass of the companion star, and the mass of the compact object.
Thus, given the mass range of the companion star, we can decide the range of the compact object by determining the range of the inclination angle.
We obtain that the mass of the compact object is 3-7 $M_{\odot}$, for $i = 15^{\circ} \text{-} 30^{\circ}$ and $M_* = 22.9^{+3.4}_{-2.9}\ M_{\odot}$ 
\citep{cas05}.
This means that the compact object is a black hole under the assumption of the constant injection.

If we do not assume the constant injection, the observational data might be reproduced even for $i > 30^{\circ}$, which would allow the lower 
limit on the mass of the compact object under 3 $M_{\odot}$. However, this requires the assumption that the injection rate is smaller around SUPC phase,
which is near the periastron, and that it is larger around INFC phase, which is near the apastron, which is contrary to naive expectation.

\section{Summary}
\label{sum}
We calculate X-ray, GeV, and TeV spectra and light curves from LS5039 with the same band as observed by HESS, Fermi and Suzaku, respectively 
under three kinds of assumptions.
First, electrons are injected constantly and isotropically at the location of the compact object. Second, these electrons lose energy only
by IC cooling against stellar photons.
Therefore, electromagnetic cascade develops as long as electrons have energy larger than $\sim 30\text{ GeV}$, because they radiate
IC photons which also produce $e^{\pm}$ pairs.
Finally, these electrons and positrons radiate at the position of injection or production.
Last two assumptions set limits on the uniform magnetic field in the system, around $B \sim 0.1 \text{ G}$.

\begin{figure}
\begin{center}
\includegraphics[height=6cm,clip]{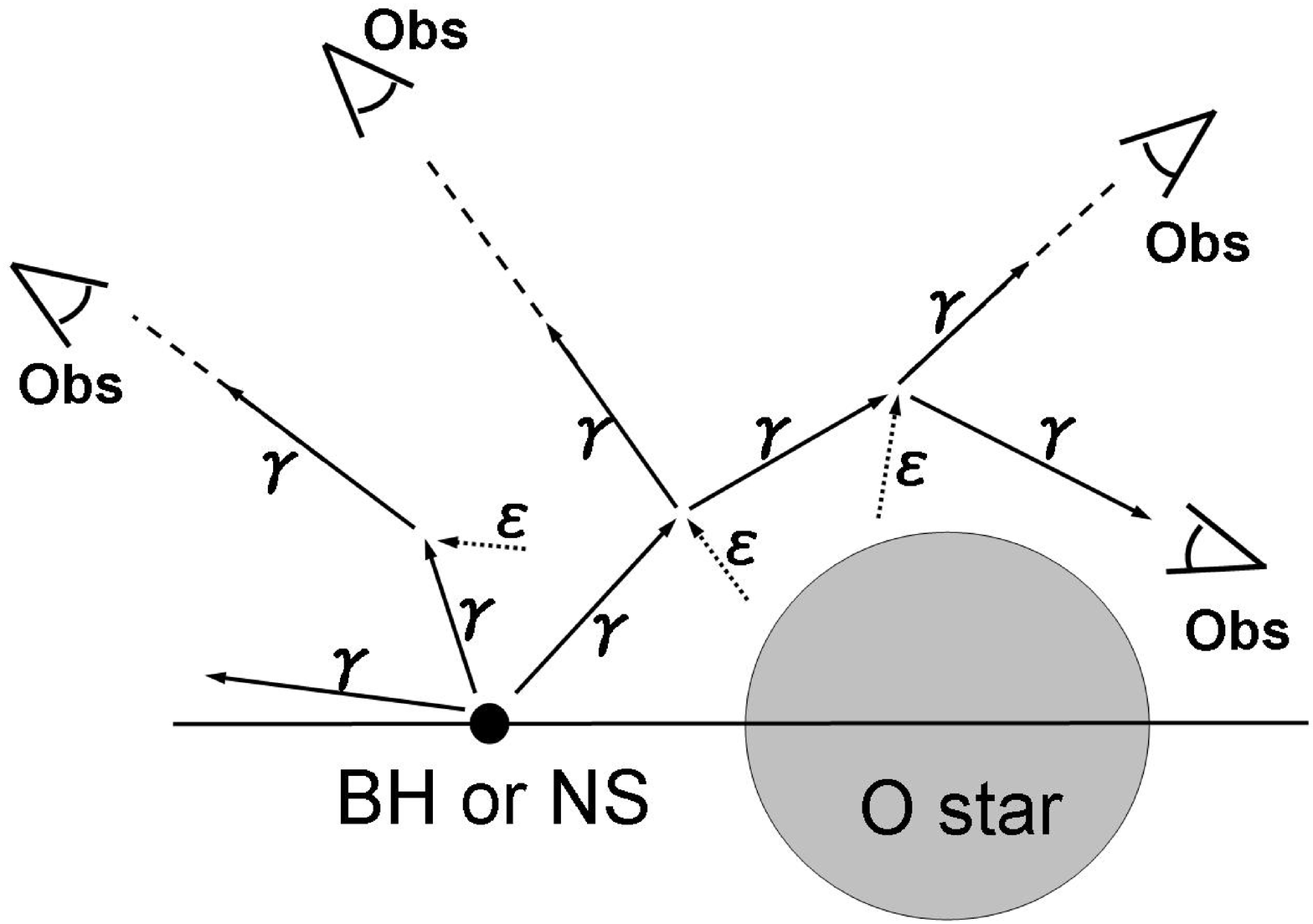}
\includegraphics[height=6cm,clip]{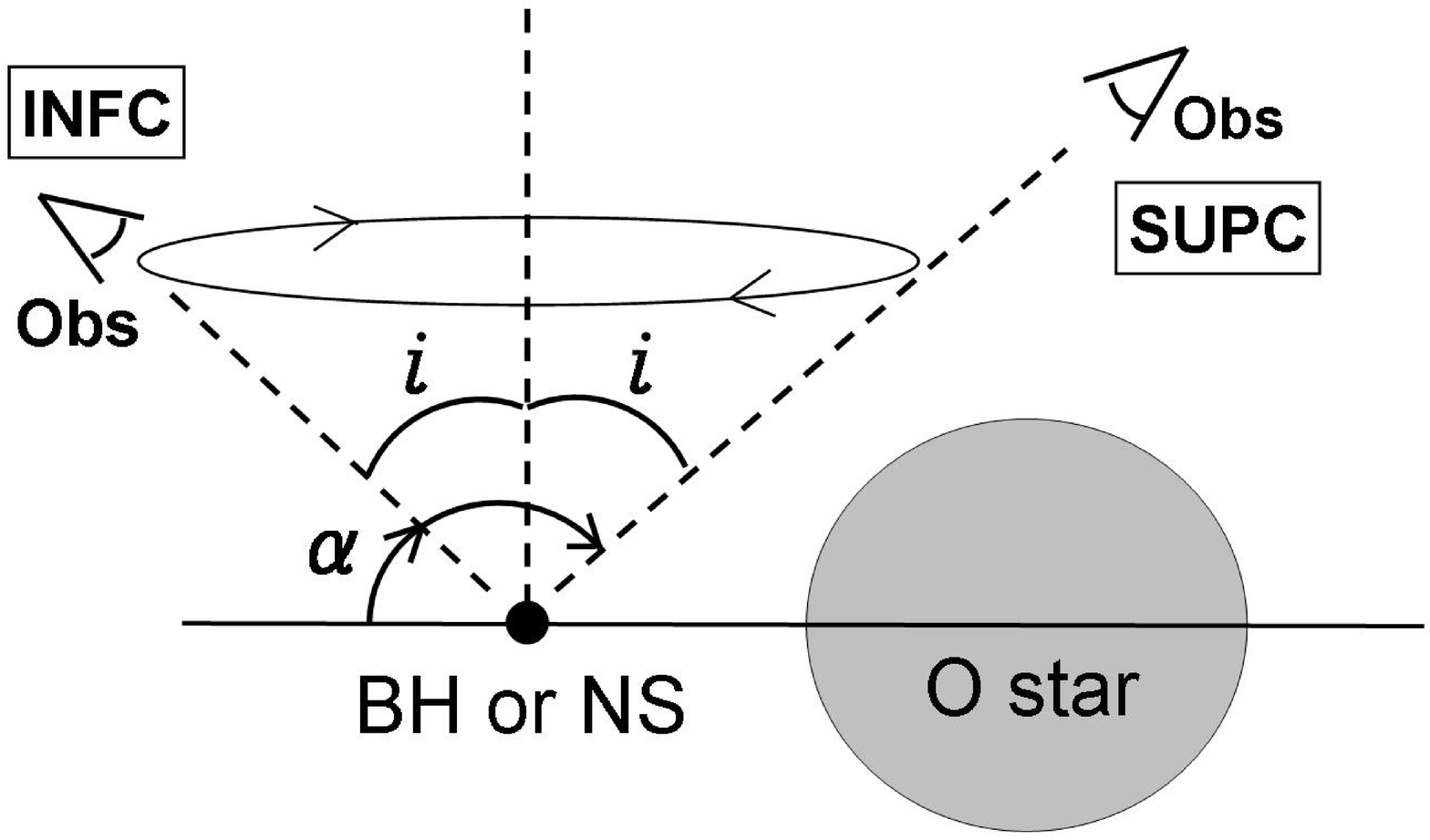}
\caption{Model pictures of LS5039 showing photon propagation (the left picture) and orbital geometry (the right picture).
LS5039 consists of a compact object, a black hole (BH) or a neutron star (NS), and an companion O star. 
$\gamma$ and $\epsilon$ represent a high energy photon and thermal photon from the companion, respectively. 
The solid line joining the center of the compact object and the companion represents a binary axis.
INFC and SUPC mean inferior conjunction, where the compact object is located between the observer and the O star, and
superior conjunction, where it is located behind the O star with respect to the observer, respectively.
$\alpha$ represents the angle between the binary axis and the line of sight (or the path of photon leaving the compact object). 
The larger $\alpha$ becomes, the thicker the radiation field by the companion becomes.
We can count escaped photons in any direction which undergo cascade process by calculating the photon propagation in 3D coordinate system.
Fixing an inclination angle $i$ and rotating the line of sight around the line perpendicular to the binary axis, 
we can obtain the phase-divided spectra and light curves for the inclination angle.}
\label{modpic}
\end{center}
\end{figure}

\begin{figure}
\begin{center}
\includegraphics[height=8cm,angle=270,clip]{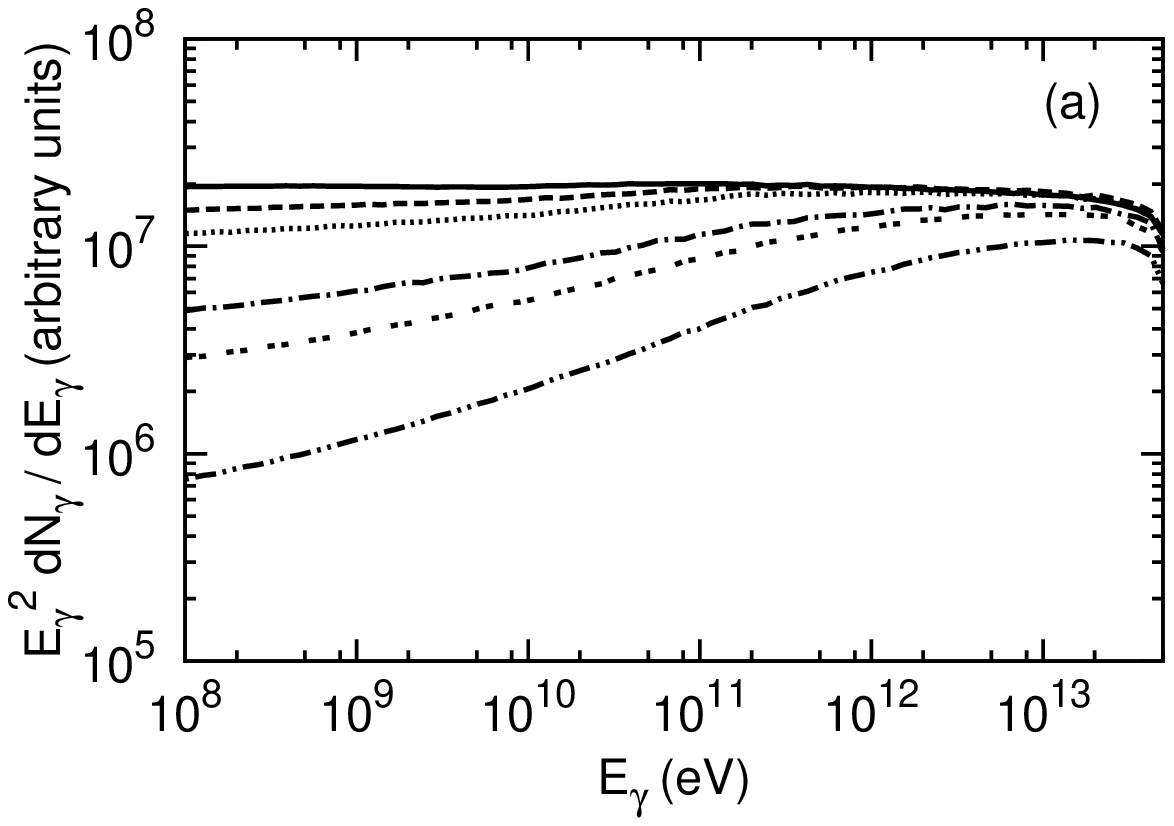}
\includegraphics[height=8cm,angle=270,clip]{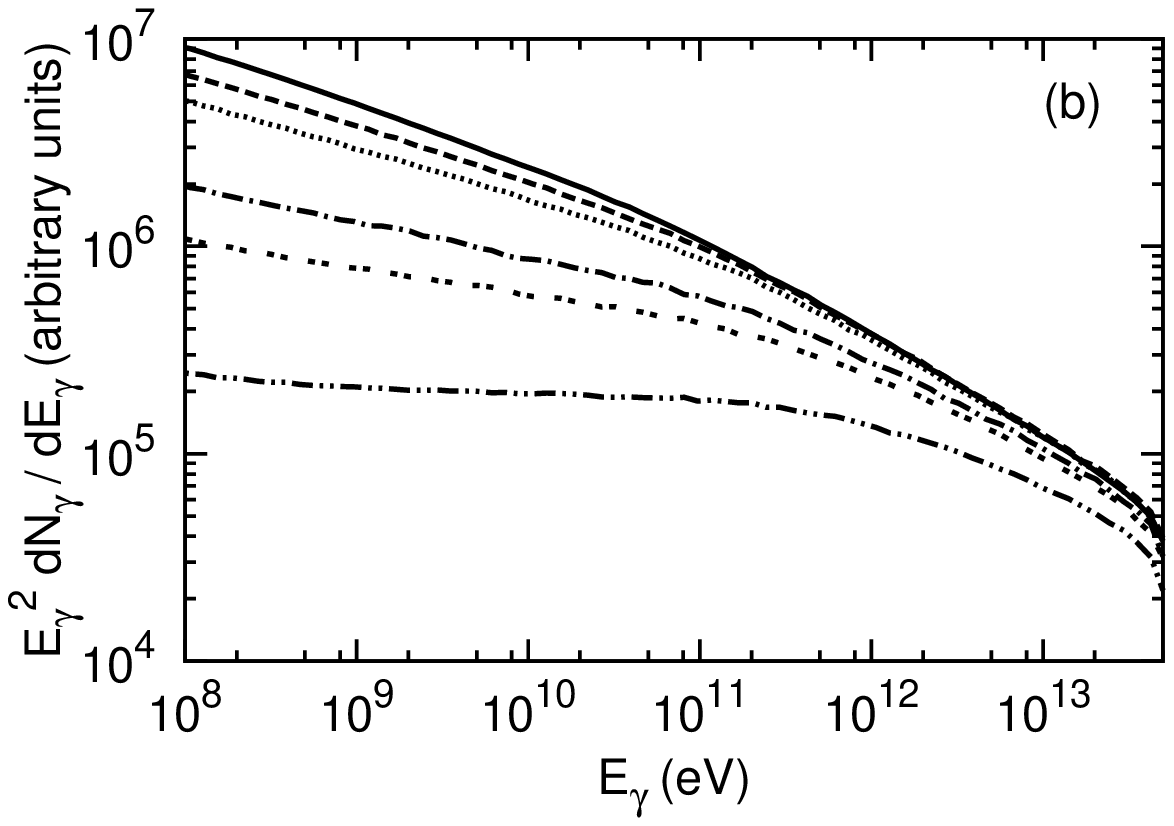}
\caption{Anisotropic inverse Compton spectrum. The seed photons for IC process come from the companion star
with a black body spectrum ($k_B T=3.3$eV), and injected electrons have a power-law distribution, namely $p = 2.0$ (a) and 
$p = 2.5$ (b). The spectra are calculated with angle $\alpha = 150^{\circ}$(solid line), $120^{\circ}$(dashed),
$105^{\circ}$(dotted), $75^{\circ}$(dot-dashed), $60^{\circ}$(double-dotted), and $30^{\circ}$(double-dot-dashed), where $\alpha$ is 
the angle between photon path and the binary axis (Figure \ref{modpic}).}
\label{noabs}
\end{center}
\end{figure}

By performing Monte Carlo simulations, we have shown that the observed spectra and light curves can be reproduced qualitatively.
The numerical calculations with these assumptions yield two important results. One is that variation of GeV flux synchronized with orbital period 
depends almost only on the change in $\alpha$ (Figure \ref{modpic}).
It is hardly influenced by the change in the separation, that is, the change in thickness of radiation field and 
the variation in the degree of cascade. Thus, we can determine the range of the inclination angle from variation of GeV flux
since the change in $\alpha$ is determined by the inclination angle.
The range of the inclination angle in turn determines mass of the compact object as 3-7 $M_{\odot}$, which implies the compact object in LS5039
is a black hole.
The other is that X-ray flux varies with binary separation, because the number density of electrons and positrons varies with thickness of
the thermal photon field by IC cooling process.

From these results and the fact that the TeV flux varies with $\alpha$ by the variation of the optical depth, 
we can derive TeV-GeV anticorrelation and TeV-X correlation, namely, TeV flux anticorrelates with $\alpha$ but GeV flux correlates with it, 
so that TeV and GeV flux anticorrelate. In addition, X-ray flux correlates with the separation, so that TeV and X-ray flux appear to correlate because of the orbital geometry of LS5039.

However, this model has several problems in detail in terms of comparison with the observational data. First of all, X-ray flux is drastically 
underpredicted for $B = 0.1$ G,
though the photon index is reproduced well. Second, the flux in 1-10 GeV range is overpredicted, in other words, the cutoff energy is overpredicted.
Third, photon index around SUPC at photon energy $\gtrsim$ 1 TeV is smaller than 2, while that of the observational data is larger than 2. 
Finally, X-ray flux modulation shows an advance by $\Delta \phi \sim 0.15$, compared to the observation data.
These problems are deferred to future research.

The most important assumption which must be modified is cooling process. X-ray flux and TeV flux at SUPC and INFC would be modified by taking into account
synchrotron cooling, though the efficient acceleration must be assumed as stated Section \ref{comp}. 

\acknowledgments
M.S.Y. acknowledges Y. Ohira, S. J. Tanaka and
the anonymous referee for useful comments.
This work is supported by KAKENHI (F. T.; 20540231).

\clearpage



\clearpage

\end{document}